\documentclass[12pt]{article}
\usepackage{amsmath}
\usepackage{amssymb}
\usepackage{amscd}
\usepackage{graphicx}
\newtheorem{theor}{Theorem}[section]
\newtheorem{lem}[theor]{Lemma}

\newtheorem{rmq}[theor]{Remark}

\let\BBox\Box
\def\Box{$\BBox$}

\def\N{\hbox{\bb N}}
\def\Z{\hbox{\bb Z}}
\def\Z{\mbox{Z\hspace{-0.3em}Z}}
\def\N{\mbox{I\hspace{-.15em}N}}

\newcommand{\Nn}{\tilde{\mathbb{N}}}
\def\R{\mbox{I\hspace{-.15em}R}}
\newcommand{\Q}{\mathbb{Q}}
\def\b{\begin{equation}}
\def\e{\end{equation}}
\def\bd{\begin{displaystyle}}
\def\ed{\end{displaystyle}}
\def\ba{\begin{eqnarray}}
\def\ea{\end{eqnarray}}
\newtheorem{theorem}{Theorem}

\newtheorem{definition}[theorem]{Definition}

\newtheorem{lemma}[theorem]{Lemma}

\newtheorem{proposition}[theorem]{Proposition}
\newtheorem{remark}[theorem]{Remark}

\newenvironment{proof}[1][Proof]{\textbf{#1.} }{\ \rule{0.5em}{0.5em}}

\def\beaa{\begin{eqnarray*}}
\def\eeaa{\end{eqnarray*}}

\newcommand{\Zz}{\tilde{\mathbb{Z}}}

\begin{document} 
\title{Multidimensional
Gaussian sums arising from  distribution of Birkhoff sums in  
zero entropy dynamical systems} 

\author{M. Bernardo$^{1, 2}$, M.Courbage$^1$\thanks{ corresponding author. e-mail:courbage@ccr.jussieu.fr,
Tel. 33-1 44 27 57 41,Fax: 33-1 46 33 94 01} and T.T.Truong $^2$}
\date{\today}      
\maketitle
\centerline{\it 1) Universit\'e Paris 7 - Denis Diderot / L.P.T.M.C.}
\centerline{\it Tour 24-14.5\`eme \'etage, 4, Place Jussieu}
\centerline{\it 75251 Paris Cedex 05, France}
\centerline{\it 2) Laboratoire de Physique Th\'eorique et Mod\'elisation}
\centerline{\it Universit\'e de Cergy-Pontoise, }
\centerline{\it F-95031 Cergy-Pontoise Cedex, France}

\begin{abstract}
A duality formula, of the  Hardy and Littlewood type
for multidimensional Gaussian sums,  is proved in    
order to estimate the asymptotic long time behavior of  distribution of Birkhoff 
sums $S_n$ of a sequence generated by a 
skew product dynamical system on the $\mathbb{T}^2$ torus, with zero Lyapounov exponents. 
The sequence, taking the values $\pm
1$,  is pairwise
independent (but not independent) ergodic sequence  with infinite range dependence.  The model 
 corresponds to the motion  
 of a particle on an infinite cylinder, hopping backward and forward along its axis, with a transversal acceleration parameter
$\alpha$. We show that when the  
parameter $\alpha /\pi$ is rational then all the moments of the 
normalized sums 
$E\left( (S_n/\sqrt{n}\right)^k)$, but  the second,  are unbounded with respect to n, while for
irrational $\alpha /\pi$, with bounded continuous  fraction representation,  all these moments  are finite and bounded with
respect to n. 

\end{abstract}

PACS numbers 05.40; 05.45

\newpage
\pagestyle{headings}

\section{Introduction}
\setcounter{equation}{0}
\indent
 
Diffusion processes are  the most important transport processes to be modelled in the frame of dynamical systems. For
a particle having deterministic "chaotic" motion, diffusion appears in the  long-time limit of  the displacement
 governed 
by the  dynamics under invariant probability measure. The  extension of the
Central Limit Theorem to deterministic dynamical systems is well established for many strongly chaotic systems. In these
systems diffusion is related to strong mixing properties of the motion of the  particles. 
However, this condition is sufficient, but
not necessary.

A stationary bounded sequence of identically distributed random variables $X_{0}, X_{1},  \ldots, $ with zero
mean value is ergodic if: 

\begin{equation}
\frac{1}{n}\sum_{i=0}^{n-1} X_{i} \underset{n \to \infty}{\longmapsto} E(X_{0})=0
\end{equation}
It fulfills the Central Limit Theorem if there exists a positive
sequence $ f_n$  such that $\frac{1}{f_n}\sum_{i=0}^{n-1} X_{i} $
converges in law to a normal centered distribution with variance
1. The validity of the Central Limit Theorem goes beyond
the case of independent (Bernoulli)  sequences  $X_{i}$. Recently this problem
has been cast in the more general setting of ergodic dynamical systems.

Let $T$ be a measurable transformation on a space $\mathcal{X}$,
equipped with a probability measure $\mu$ on a $\sigma$-algebra
$\mathcal F$ of measurable sets. The measure $\mu$ is supposed to
be $T$-invariant: for all $A$ of $\mathcal F$ one has
$\mu(T^{-1}A)=\mu(A)$. The system is also assumed to be ergodic:
\begin{equation}
\forall  f  \in {\mathcal L}^{1}(\mathcal{X});  \quad   \frac{1}{n}\sum_{i=0}^{n-1} f(T^{n}X)
\underset{n \to \infty}{\longmapsto}\int_{X}f(x)d\mu(x)=E(f)
\end{equation}
almost everywhere. The sum: $$ S_nf(X) =  \sum_{i=0}^{n-1} f(T^{n}X)$$ is called a {\bf Birkhoff sum}. A real square-integrable function $f$, with zero mean value, is said to satisfy the Central Limit Theorem if:
\begin{equation}
\lim_{n \to \infty}\mu\left(X:\frac{1}{\sqrt{<S_n^2>}}\sum_{i=0}^{n-1} f(T^{n}X)  \in I\right) =
\frac{1}{\sqrt{2\pi}}\int_{I}\exp(-\frac{x^{2}}{2})dx
\end{equation}
In other words, the sum $\frac{1}{\sqrt{(<S_nf>)^2}}\sum_{i=0}^{n-1} f\circ
T^{n}$ converges in law to a Gaussian variable of zero mean value
and variance 1.
A  generalization of the Central Limit Theorem has been
established for a class of regular functions for K-systems [Go], and for hyperbolic dynamical systems the
auto-correlations of which decrease exponentially (see a review in [Li]). One of the most remarkable results of the theory of chaotic dynamical systems was the convergence of
normalized Birkhoff sums of  observables to a diffusion process. In the Lorentz gas, Bunimovich, Chernov, Sinai [BCS],
[BS]
 and Young [CYo] have proved such behavior.  They found a class of functions $f$
for wich the rescaled ergodic sums converge to the Wiener process, i.e.
\begin{eqnarray}
\sqrt{\tau}\sum_{n=0}^{[t/ \tau]} f\circ T^n \underset{\tau \to 0}{\to} Y_t
\end{eqnarray}
where $t\in[0,1]$ and $Y_t$ is a Brownian motion (other informations can be found in [Bi]). In non integrable area preserving
maps numerical results on diffusion have been obtained by many authors  (see, for instance,  references in [Za], [MMP]).
However to our knowledge, in the hamiltonian dynamical systems no rigourous proofs have been given. In [Za], there are
several numerical evidences of the existence of anomalous transport in hamiltonian non integrable systems on account of the
existence of islands into islands in the phase  space of the systems due to  the abundance of some stable periodic behaviors
generating Levy flights. One of the main motivation of our work is to investigate diffusive
behavior in simple ergodic area-preserving  mappings with weakly chaotic
properties  (i.e. randomness with  zero Lyapounov exponenets  [CH3]). In such systems, the problem is to find a class of functions $f$
for wich the rescaled ergodic sums converge to the Wiener process. A first step toward such result is to  estimate the asymptotic behavior of the Birkhoff
sums. In this work we consider this problem for completely non-hyperbolic systems having nevertheless
a family of observables for which the two-times auto-correlations are zero.
Swante Janson [J] has studied examples of nonergodic stationary sequences with
zero mean value and variance $\sigma$, such that the sequence
$S_{n}=\sum_{i=0}^{n-1}X_{i}$ converges in law, without normalization, which implies a breaking of the Central Limit Theorem. 

We shall consider an ergodic  sequence of  pairwise independent variables, each of them taking values in $\{-1,1\}$
and having variance $\sigma^{2}=1$. They are derived
from a dynamical system on the torus $\mathbb {T}^{2}$, called the Anza\"{i} skew product and defined by:
\begin{equation}
\label{2eq0}
T(x,y)\equiv (x+\alpha,y+x) \mod 2\pi
\end{equation}

When $\alpha/\pi$ is irrational, Furstenberg (see ref. in [CFS]) has shown that this
transformation has a unique invariant measure, which is thus
ergodic and is in fact the normalized Lebesgue measure $
\frac{dxdy}{(2\pi)^{2}}$. In a previous work [CH1] and [CH2]  have
determined partitions $\{ A, A^{C}\}$, which are pairwise
independent: $\mu( A\cap T^{-n}A)=\mu(A)^{2}$. It is equivalent to say
that for the function $f= \mathbb{I}_{A}-\mu(A)$  the family of functions $f\circ T^{n}$
are pairwise independent random variables.
However, this does not imply at all that these variables are jointly independant. 
Actually, the process has infinite memory because its metric entropy is zero, moreover, there is
only linear divergence of trajectories. It is to be noted that on account of the pairwise
independence, it can be immediately seen that the "diffusion
coefficient" $\sigma$ defined by: 

$$
\frac{1}{n} \int_{I}(\sum_{i=0}^{n-1} f(T^{n}x))^2dx  \underset{n \to
\infty}{\longmapsto} \sigma^2
$$
\noindent
is finite.
However,  the existence of the diffusion coefficient  is not sufficient to imply asymptotic normal
distribution.

Thus it would be interesting to see
whether the Central Limit Theorem holds here or not. We shall see
that the result depends essentially on the continued fraction
representation of $\alpha$. In this work,  $A$
will be limited to the subset $\mathbb{T} \times [0,\pi]$, $X_i=f\circ T^i$  and the
moments $E((S^{q}_{n}))$ will be asymptotically estimated. It is however difficult to 
reconstruct the characteristic function $\Phi_{n}(t)$ of $\frac{S_{n}}{\sqrt{n}}$.
 For  odd $q$, $E(S_n^q)=0$. For even $q$ several situations can occur: We shall show that:

{\bf a)} If $\alpha/\pi$ is an irrational number with a bounded
continued fraction representation, then $E((S^{2q}_{n})= \mathcal{O}(n^{q})$. 
Consequently  $\mid E((\frac{S_{n}}{\surd n})^{2q})\mid \leq A_{q}$, where $A_{q}$ is a constant. The convergence of the
characteristic function depends on the $A_{q}$.

{\bf b)} If $\alpha/\pi$ is an irrational but not of the previous type, then there is no
$f_n$  such that $E((\frac{S_{n}}{f_n})^{2k})=\mathcal{
O}(1) $, different from zero for all $k\geq 0$. \\

It is to be noted that our mapping gives interesting examples where the existence of th diffusion coefficient
$\sigma^2=\lim_{n \to _\infty}E(S_n^2)/n=1$ is not sufficient to imply the convergence to the normal distribution.

So it may be suggestive to start out with a concrete
picture of particle dynamics. The dynamical system Eq. \ref{2eq0} corresponds to the motion of a rotator kicked at regular time interval by a force modulated so that the angular velocity remains bounded.
Let $\theta(t)$ and $\dot{\theta}(t)$ be the angle and the angular velocity of the rotator, where $\dot{\theta}(t) \in [0,2\pi[$.
Then, at time $t+1$, the state of the rotator is given by
\begin{eqnarray}
\theta(t+1) &\equiv &  \theta(t)+\dot{\theta}(t) \mod (2\pi) \nonumber\\
\dot{\theta}(t+1) & \equiv & \dot{\theta}(t)+\alpha \mod (2\pi)
\end{eqnarray}
and we have at time $t$:
\begin{eqnarray}
\theta(t) \equiv   \theta(0)+t \dot{\theta}(0)+\frac{t(t-1)\alpha}{2} \mod (2\pi)
\end{eqnarray}
An idea of the randomness of this motion can be seen by using a partition $\{ A, A^{C}\}$ of the torus into two regions: $ A = \{ \theta \in  [0,
\pi [\}$ and
$ A^{C} =
\{ 
\theta
\in  [\pi, 2\pi [\}$. It can be seen  that: 

$$
\mu(A_i\cap T^{-n}A_j)  = \mu(A_i) \mu(A_j) $$  
for any $n \neq 0 $, where $A_i$ is either $A$ or $A^c$ [CH1].  This partition is far from being a Bernoulli one since the entropy of the system
is zero. We call such partitions pairwise independent. In [CH3], it is shown that such sequences are unpredictable in the
 sense of Wiener least 
squares  criterion.  It is natural to study the distributions of sums of such sequences, which  represent  a  
particle displacement induced by the dynamical system of the Eq. (\ref{2eq0}) as follows. 
\\ 

A particle moves on an infinite plane among periodically distributed obstacles with spatial period equal to 1 along both
$(q_1,q_2)$-directions.  In the 
$q_1$-direction, the motion of the particle is uniformly accelerated at each regular time interval  by an amount $\alpha
$   and has uniform free motion
along the $q_2$-direction. That is, define the velocity $p_1(n) = q_1(n+1) - q_1(n)$ , then the equations of the
projection of the  motion in the 
$q_1$-direction  are :
\ba
q_1(n) = q_1(n-1) + p_1(n-1) \\
p_1(n)  =  p_1(n-1) + \alpha
\ea
It is the result of the discrete time  action of the mapping: $T: (p_1,q_1) \rightarrow T(p_1,q_1) $ given by:
\ba
T(p_1,q_1) = (p_1+\alpha, q_1+p_1)
\ea
Thus,  we obtain:
\ba
q_1(n) = q_1(0) + np_1(0) + n(n-1)\alpha/2
\ea
The particle is moreover submitted at the begining of each time interval to a field changing
   the direction of the motion up and down along $ q_2$-direction in terms of its position along the $
q_1$-direction in the following way (see figure 1): the
velocity direction of the particle  $p_2$  at time $t=n$ is given by  $\chi(q_1(n))$ where $\chi(x)$ is a 
periodic discontinuous function defined by:

\begin{equation}
\chi(x ) =\left \lbrace
\begin{array}{ccc}
-1  & if  & x\in ]0,1/2] \\
1 & if & x\in ]1/2,1]
\end{array}
\right.
\end{equation}

\begin{figure}

\includegraphics[width=\columnwidth]{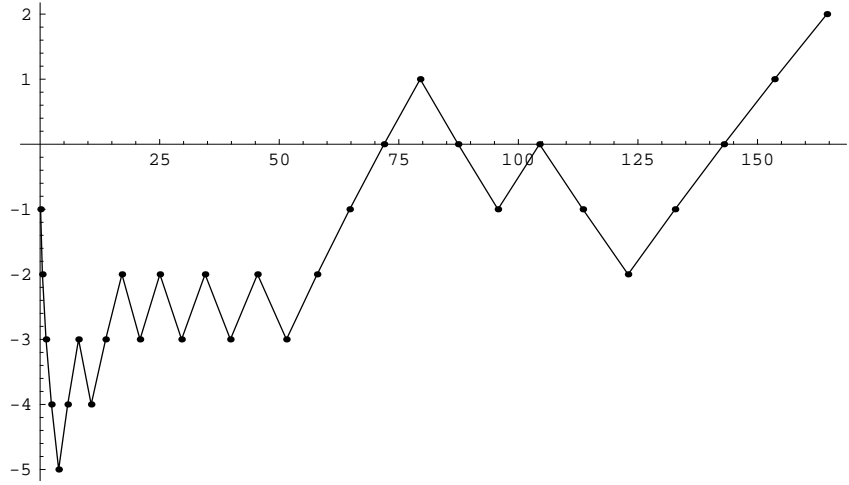}

\caption{Trajectory of the particle in the plane $ (q_1, q_2) $.}

\end{figure}

That is, the direction of the velocity at time $t=n$ is : 
\ba
\varepsilon_n (p_1(0), q_1(0))  = \chi(q_1(n)) = \chi( q_1(0) + np_1(0) + n(n-1)\alpha/2) 
\ea
and  the value of the variable
$q_2$ at time
$t = n+1$ is:
\begin{eqnarray}
q_2(n+1)& = &q_2(n) +\chi(q_1(n)) = \sum_1^n \varepsilon_i (p_1(0), q_1(0))
\end{eqnarray}

 Let us
denote by $S_k$ the position  of the particle along the $q_2$--direction at $t = k$.  The equation for such a
motion is expressed by the following recursion relation:
\begin{eqnarray}
S_{n+1}& = &S_n +\chi(q_1(n)) . \nonumber\\
S_0 & = & 0
\end{eqnarray}

In terms of the variables $(x,y) \in  \mathbb{T}^2$ the distance travelled by the particle along the vertical direction 
after $n$ steps  is:
\begin{equation}
\label{2eq1}
S_n(x,y;\alpha)=\sum_{k=0}^{n-1}\chi(y(k)) =\sum_{k=0}^{n-1}\chi
\left( y+kx+\frac{k(k-1)}{2}\alpha \right)
\end{equation}

The problem is to study the limiting distribution of such random walk. \\

In the following sections, the proof will be given in steps with
an increasing order of complexity, which turns out to be related to
number theoretical questions. The main objective is to obtain the
asymptotic behavior of the expectation values  of 
$S_{n}^{q}(\alpha)$ for $n\longrightarrow \infty$. In the course of this study, we encounter the so-called
 multidimensional Gaussian
sums which are a particular case of the Weyl sums [K]. They
are essentially sums of the form used in the definition of theta
functions, except that the summation is finite. 
 The Gaussian sums
fulfill an exact duality formula, known as Landsberg-Schaar
formula [L], which is a variant of the famous Jacobi identity for
theta functions. Except for the rational assumptions in the
application of Schaar formula, some works had been done to
estimate the growth rate of the one dimensional Gaussian sums  by Hardy and
Littlewood in [HL].  To this end they have first established a duality formula and then  used the theory of continuous
fraction decomposition (here we use tools and notations from [VdP] and [R]). We shall treat
respectively in sections 3, 4 and 5 Gaussian sums in one, two and
$d$ dimensions, the central  point in each part is concentrated in
the proof of a duality formula for these sums.

The Gaussian sums have been more
recently studied from a geometrical point of view. [DM] and [Z] have
applied the results of [HL] on polynomials associated to modular
functions. Interestingly one may derive this identity from a
simple quantum mechanical system having the torus as phase space 
[AR]. But Gaussian sums are found also in other fields like the problem of
fractional wave packet revival [SLB].
$\\$

\section[The rational case]{Preliminary considerations and the case of $\alpha$ rational}
\setcounter{equation}{0}
Before going into the crux of the
subject, let us introduce some technical steps. The important
quantity to study here is the $q\textsuperscript{th}$ moment
defined by:
\begin{equation}
E(S_n^q(\alpha))=\int_{[0,2\pi [ \times
[0,2\pi[}(S_n^q(x,y;\alpha))\frac{dx.dy}{4\pi^2}
\end{equation}
An objectve is to study the convergence of the characteristic function:
\begin{equation}
\Phi_n(t)=\int_{[0,2\pi [ \times
[0,2\pi[}e^{it.(S_n(x,y;\alpha)/\sqrt{n})}\frac{dx.dy}{4\pi^2}=\sum_{q=0}^\infty
\frac{(it)^q}{q!} E(S_n^q(\alpha)/\sqrt{n})
\end{equation}
Using the Fourier decomposition of the function $\chi$:
\begin{equation}
\chi(y)=\frac{2i}{\pi}\sum_{p\in \Zz}\frac{e^{iPy}}{P}
\end{equation}
where $\Zz$ is the set of relative odd integers, we have an
alternative form of the distance travelled by the particle:
\begin{equation}
S_n(x,y;\alpha)=\frac{2i}{\pi}\sum_{P\in
\Zz}\sum_{k=0}^{n-1}\frac{e^{iP\left(y+kx+\frac{k(k-1)}{2}\alpha
\right)}}{P}.
\end{equation}

Hence with this form one may compute the $q\textsuperscript{th}$
moment. After integration on $x$ and $y$, which yields two  
constraints:
\begin{eqnarray}
\sum_{i=1}^q P_i & = & 0 \nonumber \\
\sum_{i=1}^q k_iP_i & = & 0
\end{eqnarray}

 The above expression becomes:
\begin{equation}
E[S_n^q( \alpha)]=\left( \frac{2i}{\pi} \right) ^q \sum_{P_1\in
\Zz}\ldots  \sum_{P_q\in \Zz}\sum_{k_1=0}^{n-1}\ldots
\sum_{k_q=0}^{n-1}\frac{1}{P_1\ldots P_q} e^{i 2\pi\beta(P_1
k_1^2+\ldots+P_qk_q^2)}
\end{equation}

where  $\beta=\frac{\alpha}{4\pi}$. We are now in a position to prove:

\begin{lem}
\label{lem1}
\begin{eqnarray}
\label{eq3207}
E[S_n^{2q+1}(\alpha)] & = & 0 \nonumber \\
E[S_n^{2q}(\alpha)] & = & (-1)^q \frac{4^q}{\pi^{2q}}
\sum_{P_1\in \Zz}\ldots  \sum_{P_2q\in \Zz}\sum_{k_1=0}^{n-1}
            \ldots \sum_{k_2q=0}^{n-1}\frac{ \delta\left(
\sum_{i=2}^{2q}P_i(k_i-k_1)\right)}{P_1\ldots P_2q} \nonumber \\
            & & \times \exp{i\pi \beta\sum_{i=2}^{2q} \sum_{j=2}^{2q} \frac{P_iP_j}{\sum_{l=2}^{2q} P_l}(k_i-k_j)^2}
\end{eqnarray}
\end{lem}
where $\delta(.)$ is the kronecker function: equal to $1$ when the argument is $0$ and 0 otherwise.
$\\$

\begin{proof}
$(P_i)_{i=1,\ldots,2q}$  are odd. The condition  $\sum_{i=1}^q
P_i=0$ can only be satisfied if $q$ is even. Hence
$E[S_n^{2q+1}(\alpha)]=0$. Now from the constraint
$\sum_{i=1}^{2q}P_i=0$, we get $P_1=-\sum_{i=2}^{2q}P_i$ and from
$\sum_{i=1}^{2q}k_iP_i=0$ i.e. $ \sum_{i=2}^2q
P_i(k_i-k_1)=0$. Thus, working out $k_1$:
\begin{equation}
k_1=\frac{ \sum_{i=2}^{2q}P_ik_i }{\sum_{i=2}^{2q}P_i }
\end{equation}
with $(k_i)_{i=2,\ldots,n-1} \in \{0,\ldots,n-1\}$. And we obtain, by substitution:
\begin{eqnarray}
\label{eq3209}
\sum_{i=1}^{2q}P_ik_i^2 & = & \frac{ 1}{\sum_{i=2}^{2q}P_i }\left( \frac{1}{2}\sum_{i=2}^{2q} \sum_{j=2}^{2q} P_iP_j(k_i^2+k_j^2)-\sum_{i=2}^{2q} \sum_{j=2}^{2q}P_iP_jk_ik_j \right) \nonumber \\
              & = & \frac{ 1}{\sum_{i=2}^{2q}P_i }\left( \frac{1}{2}\sum_{i=2}^{2q} \sum_{j=2}^{2q} P_iP_j(k_i-k_j)^2 \right)
\end{eqnarray}
which ends the proof.\end{proof}\\

In particular $E[S_n^2(\alpha)]$ can be exactly and directly computed :
\begin{equation}
E[S_n^2(\alpha)]=\frac{4}{\pi^2} \sum_{P_1 \in \Zz}  \sum_{P_2 \in
\Zz} \sum_{k_1=0}^{n-1} \sum_{k_2=0}^{n-1} \frac{e^{-i
2 \pi \beta (P_1k_1^2+P_2k_2^2)}}{P_1P_2}
\end{equation}
But we get from the constraints, $P_1=-P_2=P$ and $k_1=k_2=k$. One
has thus:
\begin{equation}
E[S_n^2(\alpha)]=\frac{4}{\pi^2} \sum_{P\in \Zz} \frac{1}{P^2}
\sum_{k=0}^{n-1} 1
\end{equation}
From the property of the Riemann Zeta-function [MOS],
it may be shown that $\sum_{P \in \Zz} 1/P^2=\pi^2/4$. Hence
\begin{equation}
E[S_n^2({ \alpha})]=n
\end{equation}

The fact that odd moments are zero implies that the
characteristic function is also even and consequently the Fourier
inverse transformation of the characteristic function, that is the
probability distribution function of the random variable
$S_n(x,y;\alpha)$ is also even.

Finally we observe that the series
\begin{equation}
\sum_{(P_2,\ldots,P_{2q}) \in \Zz^{2q-1}} \frac{1}{P_2\ldots
P_{2q}(P_2+\ldots P_{2q})}
\end{equation}
is absolutely convergent since, for all $(P_2,\ldots,P_{2q}) \in
\Zz^{2q-1}$, $\vert P_2+ \ldots + P_{2q} \vert \geq 1$. Let us
prove it for $q=2$. The generalization to arbitrary $q$ is straighforward.
The series may be split into:
\begin{eqnarray}
\sum_{(P_2,P_3,P_4) \in\Zz^3} \frac{1}{\vert P_2P_3P_4(P_2+P_3+P_4) \vert} & = & 2
\sum_{(P_2,P_3,P_4) \in\Nn^3} \frac{1}{\vert P_2P_3P_4(P_2+P_3+P_4) \vert} \nonumber\\
& + & 6  \sum_{(P_2,P_3,P_4) \in\Nn^3} \frac{1}{\vert P_2P_3P_4(P_2+P_3-P_4) \vert} \nonumber \\
\label{2eq1}
\end{eqnarray}
Let $(x,y,z) \in  [0,\infty[^3$ and $0<\delta P <1$ such that:

\begin{eqnarray}
1\leq P_2 \leq x < P_2+\delta P & \Leftrightarrow & x-\delta P<P_2 \leq x \nonumber \\
1\leq P_3 \leq y < P_3+\delta P & \Leftrightarrow & y-\delta P<P_3 \leq y \nonumber \\
1\leq P_4 \leq z < P_4+\delta P & \Leftrightarrow & z-\delta P<P_4
\leq z
\end{eqnarray}
The first sum in the \emph{r.h.s.} of the equation \ref{2eq1} is bounded :
\begin{eqnarray}
\sum_{(P_2,P_3,P_4) \in\Nn^3} \frac{1}{\vert P_2P_3P_4(P_2+P_3+P_4) \vert} &
\leq & \frac{1}{6^{1/3}} \sum_{(P_2,P_3,P_4) \in\Nn^3} \frac{1}{\vert P_2P_3P_4(P_2P_3P_4)^{1/ 3}\vert} \nonumber\\
&  =  &  \frac{1}{6^{1/3}} \left(  \sum_{(P) \in\Nn}
\frac{1}{P^{1+1/3}} \right)^3 < \infty
\end{eqnarray}
The second sum can be transformed using: $0< (x-\delta P)
(y-\delta P) (z-\delta P) \leq P_2P_3P_4 \leq xyz $ and $x+y-z
-2\delta P \leq P_2+P_3-P_4 \leq x+y-z+\delta P$. Hence
\begin{eqnarray}
 \frac{1}{ \vert P_2P_3P_4(P_2+P_3-P_4) \vert} & \leq &  \frac{1}{(x-\delta P) (y-\delta P)
(z-\delta P) }  \\
& &  \max \left( \frac{1}{\vert x+y-z -2\delta P \vert },  \frac{1}{\vert
x+y-z+\delta P \vert } \right)  \nonumber \\
 & \overset{def}{=} &  F_{\delta P}(x,y,z)\nonumber
\end{eqnarray}
But $\forall \, \delta P < 1$, $\exists \epsilon >0$, such  that
any hypercube $$[(P_2,P_3,P_4); \ldots ;(P_2+\delta P, P_3+\delta
P,P_4+\delta P)]$$ does not lie in the intersection $
\Omega_\epsilon \overset{def}{=}  [0,\infty[^3 \cap \{ \vert x+y-z\vert >
\epsilon \}$. Then we have
\begin{eqnarray}
\sum_{(P_2,P_3,P_4) \in\Nn^3} & &  \frac{1}{\vert P_2P_3P_4(P_2+P_3-P_4) \vert }  \\
& \leq & \frac{1}{(\delta P)^3}\sum_{(P_2,P_3,P_4) \in\Nn^3} \int_{P_2}^{P_2+\delta P} \int_{P_3}^{P_3+\delta P} \int_{P_4}^{P_4+\delta P} F_{\delta P}(x,y,z) \nonumber\\
& \leq & \frac{1}{(\delta P)^3} \int _{ [0,\infty[^3 \setminus \Omega_\epsilon} F_{\delta P}(x,y,z) \nonumber
\end{eqnarray}
One can easily estimate the left-hand-side by realizing that the
integration of the integrals:
\begin{eqnarray}
\int _{ [0,\infty[^3 \setminus \Omega_\epsilon}
\frac{dx.dy.dz}{\vert x+y-z+\delta P \vert }\quad \mbox{and} \quad \int _{
[0,\infty[^3 \setminus \Omega_\epsilon} \frac{dx.dy.dz}{\vert
x+y-z-2\chi P \vert }
\end{eqnarray}
yields an absolutely convergent result.

\subsection{ $\alpha=0$  }
\setcounter{equation}{0}
We consider the simple case $\alpha=0$ and
show that the case with $\alpha$  rational can be brought back to
the case $\alpha=0$. First consider again:
\begin{equation}
S_n(x,y;0)=\frac{2}{i\pi}\sum_{P\in \Zz} \frac{1}{P}
\sum_{k=0}^{n-1} e^{iP(kx+y)}
\end{equation}
The above  geometric sum can be  reexpressed in terms of $ u=e^{ix}$ and $v=e^{iy}$ as:
\begin{equation}
S_n(x,y;0)=\frac{2}{i\pi}\sum_{P\in \Zz} \frac{1}{P}
z^{-\frac{n-1}{2}P} \left(
\frac{1-u^{nP}}{1-u^P}\right)\sum_{\epsilon =\pm 1} \epsilon
u^{\epsilon P \frac{n-1}{2}} v^{\epsilon P}
\end{equation}

Now, the $2k$\textsuperscript{th} moment
\begin{equation}
E[S_n^{2k}(0)]=\int_{[0,2\pi[\times[0,2\pi[} \left[ \frac{2}{i\pi}
\sum_{P\in \Zz}\frac{1}{P} \left( \sum_{k=0}^{n-1}e^{-iP(kx+y)}
\right) \right] ^{2k}\frac{dx.dy}{ 4\pi^2}
\end{equation}
may be computed as complex integrals evaluated on the the product
of the circles of unit radius and centered at $0$, i.e.
$\mathcal{C}(0,1)\times  \mathcal{C}(0,1)$. In the domain bounded
by these circles, the integrand is biholomorph and the product of
the contours is homotopical to $\mathcal{C}(0,1/2)\times
\mathcal{C}(0,1/2)$. Thus $(1-u^p)^{-1}$ can be expanded in power
series around $0$, with $u=e^{ix}$ et $v=e^{iy}$:

\begin{equation}
S_n(u,v;0)=\frac{2i}{\pi}  \sum_{P\in \Zz}u^{\frac{P}{2}} \left(
\sum_{\eta = \pm 1}\eta u^{\eta n \frac{P}{2}}\right) \left(
\sum_{m=0}^\infty u^{mP} \right) \left(  \sum_{\epsilon = \pm 1}
\frac{\epsilon}{P} u^{\epsilon  \frac{n-1}{2}P} v^{\epsilon P} \right)
\end{equation}
or
\begin{equation}
S_n(u,v;0)=\frac{2i}{\pi} \sum_{\overset{ \eta = \pm 1}{\underset{\epsilon = \pm 1}{}} }
\sum_{\overset{ m\in \N}{\underset{P\in \Nn}{}} } \frac{\eta \epsilon}{P}u^{P\left(
\frac{\epsilon +\eta}{2}n+m+\frac{1-\epsilon}{2} \right)}v^{\epsilon P}
\end{equation}
We can now compute $E[S_n^2(0)]$ with this expression:
\begin{eqnarray}
E[S_n^2(0)]& = & -\frac{4}{\pi^2} \sum_{\overset{\eta_1, \eta_2 = \pm 1}{\underset{ \epsilon_1, \epsilon_2  = \pm 1}{}}}
\sum_{\overset{ m_1, m_2\in N}{\underset{P_1, P_2\in \Nn}{}} }
\frac{\eta_1 \epsilon_1}{P_1}\frac{\eta_2 \epsilon_2}{P_2} \oint \frac{dv}{2i\pi}v^{\epsilon_1 P_1+\epsilon_2 P_2}\nonumber\\
& &  \oint \frac{du}{2i \pi} u^{P_1\left( \frac{\epsilon_1
+\eta_1}{2}n+m_1+\frac{1-\epsilon_1}{2} \right) + P_2\left(
\frac{\epsilon_2 +\eta_2}{2}n+m_2+\frac{1-\epsilon_2}{2} \right)} \nonumber\\
\label{2eq2}
\end{eqnarray}
The first integral over $v$ yields $\epsilon_1P_1+\epsilon_2
P_2=0$, one must have $\epsilon_1=-\epsilon_2$ and $P_1=P_2=P$.
The exponent of $z$ in the $2$\textsuperscript{nd} integral
becomes:

\begin{eqnarray}
P_1\left( \frac{\epsilon_1 +\eta_1}{2}n+m_1 + \frac{1-\epsilon_1}{2} \right) & +  & P_2\left( \frac{\epsilon_2 +\eta_2}{2}n+m_2+\frac{1-\epsilon_2}{2} \right) \nonumber\\
& = &P\left(\frac{\eta_1+\eta_2}{2}n+m_1+m_2+1 \right) \nonumber\\
\end{eqnarray}

The contribution to the complex integration comes only from the
vanishing of this quantity. this leads to the following:
\begin{eqnarray}
\eta_1=\eta_2 & = & -1\\
m_1+m_2 &=& n-1
\end{eqnarray}
and card$\{(m_1,m_2) \in \N^2$ tel que $m_1+m_2=n-1 \}=n$, thus:

\begin{equation}
E[S_n ^2(0)]= \sum_{\epsilon= \pm1} \sum_{p\in
\Nn}\frac{4n}{P^2\pi^2}=\frac{4n}{\pi^2}\sum_{P\in
\Zz}\frac{1}{P^2}=n
\end{equation}

This computational method may be extended  to higher moments
$E[S_n^{2k}(0)]$. We see that using Eq \ref{2eq2} to power $2k$
yields
\begin{eqnarray}
& & E[S_n(0)^{2k}] = (-1)^k \left(\frac{4}{\pi^2} \right)^k
\sum_{\eta_1,\ldots \eta_{2k} = \pm 1}  \sum_{\epsilon_1,\ldots \epsilon_{2k} = \pm 1} \sum_{m_1, \ldots m_{2k} \in \N}  \sum_{P_1, \ldots P_{2k} \in \Nn} \nonumber \\
& &\prod_{i=1}^{2k} \frac{\eta_i\epsilon_i}{P_i}   \oint
\frac{dv}{2i \pi} v^{(\sum_{i=1}^{2k}\epsilon_iP_i)}  \oint
\frac{du}{2i \pi}
u^{(\sum_{i=1}^{2k}P_i(m_i+\frac{1-\epsilon_i}{2}+\frac{\eta_i+
\epsilon_i}{2}n)}
\end{eqnarray}
We can perform the $v$-integration and get:
\begin{equation}
\sum_{i=1}^{2k} \epsilon_iP_i=0
\end{equation}
and the $u$-integration yields the following proposition:
\begin{proposition}
\label{prop1}
\begin{eqnarray}
E[S_n(0)^{2k}] & = & (-1)^k \left(\frac{4}{\pi^2} \right)^k
\sum_{\eta_1,\ldots \eta_{2k} = \pm 1}  \sum_{\epsilon_1,\ldots \epsilon_{2k} = \pm 1} \sum_{m_1, \ldots m_{2k} \in \N}  \sum_{P_1, \ldots P_{2k} \in \Nn} \nonumber \\
& &\prod_{i=1}^{2k} \frac{\eta_i\epsilon_i}{P_i}
A_n[\eta_i,\epsilon_i,P_i](2k)
\end{eqnarray}
where the coefficients $A_n$ are defined as:
\begin{eqnarray}
\label{2eq3}
A_n[\eta_i,\epsilon_i,P_i](2k)= card \lbrace (m_1\ldots,m_{2k}) \in \N^{2k} \, \vert  \sum_{i=1}^{2k}m_iPi=-\frac{1}{2}\sum_{i=1}^{2k}P_i(1+n\eta_i) \rbrace \nonumber \\
\end{eqnarray}
with the constraint $\sum_{i=1}^{2k}\epsilon_iPi=0$.
\end{proposition}

This result
is obtained by the {\it Residue Theorem}. It remains to determine
the behavior of $A_n$ as $n\to \infty$. We observe that not all
the combinations of $\epsilon_i$ are to be taken into account
since $(P_1,\ldots,P_n)\in \Nn^{2k}$. Thus the choice
$(\epsilon_1,\ldots,\epsilon_{2k})=(1,\ldots,1)$ is excluded by
the constraint. According to the usual notation , if the $(P_i)_{i =1,\ldots,r}$ do
not have common divisors, we note:
$(P_1,\ldots,P_r)=1$, then:
\begin{eqnarray}
card \lbrace (m_1,\ldots,m_{r})\in \N^{r} \, \vert
\sum_{i=1}^{r}m_iPi=n \rbrace
\underset{n\to\infty}{\simeq}\frac{1}{(r-1)!}\frac{n^{r-1}}{P_1\ldots P_r}
\end{eqnarray}
Before going on we prove the following lemma:

\begin{lemma}
\label{lem2}
Let $\sum_{i=1}^r\epsilon_iPi=0$ and  $(P_1,\ldots,P_r)=1$ is
equivalent to $(P_1,\ldots,$ $\hat{P_j},\ldots,P_r)=1$ where $P_j$,
is removed.
\end{lemma}

\proof \\ It is clear that if $(P_1,\ldots,\hat{P_i},\ldots,P_r)=1$
then $(P_1,P_r)=1$. Actually if
$(P_1,\ldots,\hat{P_j},\ldots,P_r)=1$ then $\not\exists s \in \Nn
\setminus \{1\}$ such that $\forall i\not=  j$, $P_i=sp_i$ and
thus $(P_1,\ldots, P_q)=1$.

Conversely if $(P_1,\ldots, P_q)=1$, but
$(P_1,\ldots,\hat{P_j},\ldots,P_r)=s$ with $s \in \Nn \setminus
\{1\}$, i.e. $\forall i\not=  j$, $P_i=sp_i$, then
$P_j=\sum_{i\not=j} \epsilon_i \epsilon_jP_i=s(\sum_{i\not=j}
\epsilon_i \epsilon_jp_i)$ which contradicts the fact that
$(P_1,\ldots, P_q)=1$. $\blacksquare$
$\\ \\$
Coming back to our estimate of $A_n$, put $\bar{\N}^q=\Nn^q\cap
\{(P_1,\ldots,P_q)=1\}$. If $(P_1,\ldots,P_{2k}) \in
\bar{\N}^{2k}$ such that $\sum_ {i=1}^{2k} \epsilon_iP_i=0$ then:
\begin{eqnarray}
A_n[\eta_i,\epsilon_i,P_i](2k) & \underset{n\to\infty}{\sim} & \frac{(-1)^{2k-1}}{(2k-1)!}
\frac{1}{\prod_{i=1}^{2k}P_i} \left( \frac{n.\sum_{i=1}^{2k}P_i \eta_i }{2}\right)^{2k-1} \,
\mbox{if}\, \sum_{i=1}^{2k}\eta_i P_i <0 \nonumber \\
& = & \left\lbrace \begin{array}{cc}  1 & if \, \sum_{i=1}^{2k}\eta_iP_i=0 \\ 0 & otherwise  \end{array}
\right.
 \label{2eq4}
\end{eqnarray}
with $\sum_{i=1}^{2k}\epsilon_iP_i=0$. Now from  Eq \ref{2eq4} the
asymptotic behavior for $n \to \infty$  of the
$2k$\textsuperscript{th} moment is:
\begin{eqnarray}
 E[S_n^{2k}(0)]  \underset{n\to\infty}{\sim}  2\frac{(-1)^{k}}{(2k-1)!{\pi^2}^{k}}   \sum_{P\in \Nn}\frac{1}{P^{2k}}
\sum_{\overset{ \eta_i = \pm 1}{\underset{ \epsilon_i  = \pm 1}{}} }
 \sum_{\overset{ P_i \in \bar{N}^{2k}}{} }
\prod_{i=1}^{2k}\frac{\epsilon_i \eta_i}{P_i^2} \left(
\sum_{i=1^{2k}} \eta_i P_i \right)^{2k-1} n^{2k-1}\nonumber\\
\end{eqnarray}
with   $\sum_{i=1}^{2k} \epsilon_i P_i=0$ and $\sum_{i=1}^{2k} \eta_i P_i>0$ . We have already seen that the series
\begin{eqnarray}
\sum_{\overset{ \eta_1,\ldots, \eta_{2k} = \pm 1}{\underset{  \epsilon_1, \ldots, \epsilon_ {2k}  = \pm 1}{}} }
 \sum_{\overset{ P_1,\ldots, P_{2k} \in
\bar{N}^{2k}}{\underset{ \sum_{i=1}^{2k} \epsilon_i P_i=0}{}} }
 \prod_{i=1}^{2k}\frac{\epsilon_i \eta_i}{P_i^2} \left( \sum_{_i=0}^{2k}\eta_i P_i \right)^{2k-1}
\end{eqnarray}
is an absolutely convergent series. Since $\sum_{P\in \Nn}
\frac{1}{P^q} \xrightarrow[q\to \infty]{}1$, we have thus the
theorem:
\begin{theor}
\label{theo1}
$\forall k\in \N$, $E[S_n^{2k}(0)] \underset{n\to\infty }{\sim}
C_kn^{2k-1}$ with
\begin{eqnarray}
& & C_1=1 \nonumber\\
& & C_k= 2 \frac{(-1)^{k}}{(2k-1)!{\pi^{2k}}} \sum_{\overset{ \eta_i,\epsilon_i = \pm 1}{\underset{  P_i \in \tilde{N}^{2k}}{}} }
\prod_{i=1}^{2k}\frac{\epsilon_i \eta_i}{P_i^2} \left( \sum_{i=1}^{2k} \eta_i P_i \right)^{2k-1}
 \end{eqnarray}
with $\sum_{i=0}^{2k}\epsilon_iP_i=0$ and $ \sum_{i=0}^{2k}\eta_iPi >0$.
\end{theor}

\subsection{Extension to rational $\beta$ ( $\alpha \in \mathbb{Q}$)}
\setcounter{equation}{0}

To this end, we regroup terms of $S_n(x,y;\alpha)$. $\forall
\alpha\in \Q$, $\exists k_0\in \N$ such that $\forall k<k_0$:
\begin{eqnarray}
k(k-1)\beta  &\not \equiv & 0 \mod(1) \nonumber\\
k_0(k_0-1)\beta  &\equiv & 0 \mod(1)
\end{eqnarray}
We shall set $n=qk_0$ and write $S_n(x,y;\alpha)$ grouping the terms such that:
\begin{equation}
\delta_s(x,y)= \sum_{n=0}^{q-1}\chi \left(
y+[s+nk_0]x+2s(s-1)\pi \beta \right)
\end{equation}
and
\begin{equation}
S_n(x,y;\alpha)=\sum_{s=0}^{k_0-1}\delta_s(x,y)
\end{equation}

with $0<s\leq k_0$. The sum can be partially performed after
inserting the definition of $\chi$:
\begin{eqnarray}
\lefteqn{ \delta_s(x,y)=\frac{2}{i\pi} \sum_{P\in \Zz} \frac{1}{P} e^{ -ik_0 \frac{P(P-1)}{2}x }
\left(\frac{e^{iPqk_0x}-1}{e^{iPk_0x}-1} \right) }\nonumber\\ & &
\left(e^{iP(y+x[s+\frac{k_0(q-1)}{2}]+2s(s-1)\beta)}-e^{-iP(y+x[s+\frac{k_0(q-1)}{2}]+2s(s-1)\beta)}
\right)
\end{eqnarray}
Call now $s^{\prime}=2s(s-1)$, $u=e^{ix}$, $v=e^{iy}$ and
$t=e^{2i \beta}$. Then
\begin{eqnarray}
\delta_s(x,y;\alpha)& = &\delta_s(u,v;t)= \frac{2}{i\pi}\sum_{P\in Zz} \frac{1}{P} u^{ -k_0
\frac{P(P-1)}{2} } \left(\frac{1-z^{Pqk_0}}{1-z^{Pk_0}} \right) \nonumber\\ &
&\sum_{\epsilon=\pm1}\epsilon z^{\epsilon P(s+\frac{q-1}{2}k_0)} v^{\epsilon P}t^{\epsilon
Ps^{\prime}}
\end{eqnarray}
We now expand $\frac{1}{1-u^{Pk_0}}$ as in the previous section,
and obtain consequently:
\begin{eqnarray}
\delta_s(u,v;t)= \frac{2}{i\pi} \sum_{\epsilon,\eta=\pm1}
\sum_{P\in \Zz,m\in \N} \frac{\epsilon \eta}{P}  v^{\epsilon
P}t^{\epsilon Ps^{\prime}}u^{P(\epsilon
s+k_0\frac{1-\epsilon}{2}+mk_0+qk_0\frac{\eta+\epsilon}{2})}
\end{eqnarray}
The expectation value:
\begin{eqnarray}
& & E(\delta_{s_1}(t)\ldots \delta_{s_{2k}}(t))=(-1)^q\left( \frac{4}{\pi}\right)^q  \sum_{\overset{ \eta_i,\epsilon_i=\pm 1 }{\underset { P_i \in \Nn,  m_i\in\N }{}}}  \prod_{i=1}^{2k}\frac{\eta_i\epsilon_i}{P_i} t^{(\sum_{i=1}^{2k} \epsilon_iP_is_i^{\prime})}\nonumber\\
& &\oint \frac{dv}{2i\pi v}v^{(\sum_{i=1}^{2k} \epsilon_iP_i)}
\oint \frac{du}{2i\pi u}u^{( \sum_{i=1}^{2k} P_i(\epsilon_i
s_i+k_0\frac{1-\epsilon_i}{2}+m_ik_0+qk_0\frac{\eta_i+\epsilon_i}{2})
)}\nonumber\\
\end{eqnarray}
 is not zero if $\sum_{i=1}^{2k} P_i(\epsilon_i
s_i+k_0\frac{1-\epsilon_i}{2}+m_ik_0+qk_0\frac{\eta_i+\epsilon_i}{2})=0$
and $\sum_{i=1}^{2k} \epsilon_iP_i=0$. As before, one can reduce
to the case $(P_1,\ldots,P_{2k})=1$. The first assertion is true
if $\sum_{i=0}^{2k_0}\epsilon_iP_is_i$ is divisible by $k_0$, with
$0\leq s_i<k_0$. The hypothesis on $P_i$ suggests that there
exists  a $P_{i_0}$ not divisible by $k_0$.

\begin{lemma}
\label{lem3}
Let $s_1,\ldots,s_{2k}$ such that
$\sum_{i=0}^{2k_0}\epsilon_iP_is_i$ is divisible by $k_0$ then
$\forall s\in \{1,\ldots, k_0\}$ and $s\not= s_{i_0}$, $k_0$ is
not a divisor of $\sum_{i=0}^{2k_0}\epsilon_iP_is_i$.
\end{lemma}
\proof Suppose that this assertion is not true: $\exists n$ such
that $\sum_{i=0}^{2k_0}\epsilon_iP_is_i=nk_0$ and $\exists
n^{\prime}$ such that
$\sum_{i=0}^{2k_0}\epsilon_iP_is_i=n^{\prime}k_0$. Hence
$\epsilon_{i_0}P_{i_0}(s_{i_0}-s)=(n-n^{\prime})k_0$  but $k_0$ is
not a divisor of $P_{i_0}$ and $\vert s_{i_0}-s\vert <k_0$, 
this is not possible. $\blacksquare$
$\\ \\$
Consequently $\mbox{card} \{(s_1,\ldots,s_2k)\in \{1,\ldots,k_0\}^{2k}\}$
such that $k_0$ is not a divisor of
$\sum_{i=1}^{2k}\epsilon_iP_is_i$ and the value
$(P_i)_{i=1,\ldots,2k}$ are mutually prime numbers of values less
than $k_0^{2k-1}$.

\begin{remark}
\label{rem1}
\end{remark}
 We have
\begin{eqnarray}
& & \mbox{card} \left[ (m_1,\ldots,m_{2k})\in \N^{2k} \vert \sum_{i=1}^{2k}  P_i(\epsilon_i s_i+k_0\frac{1-\epsilon_i}{2}+m_ik_0+qk_0\frac{\eta_i+\epsilon_i}{2}) =0 \right] \nonumber\\
& & \underset{q\to \infty}{\simeq} \mbox{card}[(m_1,\ldots,m_{2k})\in
\N^{2k} \vert \sum_{i=1}^{2k} m_iP_i =-\frac{q}{2}\sum_{i=1}^{2k}
\eta_iP_i]
\end{eqnarray}
with $(P_1,\ldots,P_{2k})=1$. Since
$\sum_{i=1}^{2k}\epsilon_iP_i=0$ and $t^{P
\sum_{i=1}^{2k}\epsilon_i P_i s_i^{\prime}}=t^{P
\sum_{i=1}^{2k}\epsilon_i P_i s_i^2/2}$. Following the previous
procedure we obtain for $n=qk_0$.

\begin{eqnarray}
 E[S_{qk_0}^{2k}(\alpha)] & \underset{q\to \infty}{\simeq} & \frac{2(-1)^{k}}{\pi^{2k}(2k-1)!}(qk_0)^{2k-1} \sum_{\overset{ \eta_i,\epsilon_i=\pm 1 }{\underset{ P_i \in \Nn}{}}} \prod_{i=1}^{2k}\frac{\eta_i\epsilon_i}{P_i^2}
\left( \sum_{i=1}^{2k}\eta_iP_i\right)^{2k-1} \frac{1}{k_0^{2k-1}} \nonumber\\
& & \sum_{p\in \Nn}\sum_{s_i=1}^{k_0} \frac{ t^{P
\sum_{i=1}^{2k}\epsilon_i P_i s_i^2/2}}{P^{2k}}
\end{eqnarray}
with the conditions $ \sum_{i=1}^{2k}\epsilon_i P_i =0$ ,$
\sum_{i=1}^{2k}\epsilon_i P_i s_i$ is divisible by $k_0$ ,
$(P_1,\ldots,P_{2q})=1$ and $\sum_{i=1}^{2k}\eta_iP_i\geq 0$. As
it was mentionned , for all $P\in \Nn$, we have
\begin{eqnarray}
\begin{array}{|c|}  \frac{1}{k_0^{2k-1}}\sum_{s_i=1}^{k_0} t^{P/2 \sum_{i=1}^{2k}\epsilon_i P_i s_i^2} \end{array} <1
\end{eqnarray}
Note that for $\alpha=0$ ($k_0=1$) this term is simply $1$. 
Consequently we have the theorem:
$\\$
\begin{theor}
\label{theo2}
$\forall \alpha \in \Q$, $\forall k\in \N^{\ast}$
$E[S_n^{2k}(\frac{\pi \beta}{4})] \underset{n\to \infty}{\simeq} C_kn^{2k-1}$
with $C_1=1$ and $C_k$ a constante.
\end{theor}

This result suggests a unique normalization, for which the
characteristic function of $S_n(x,y;\alpha)$ once normalized will
be independent of $n$, when $n\to \infty$:
\begin{equation}
\phi_{\frac{S_n}{n}}(t)\underset{n\to \infty}{\simeq} \sum_{k=0}^{\infty}\frac{(it)^{2k}}{(2k)!}
C_k\frac{n^{2k-1}}{n^{2k}}=1
\end{equation}
Consequently the probability distribution function $F$ of the normalized
random variable limit $S=\lim_{n\to \infty}\frac{S_n(x,y;\alpha)}{n}$, for
$\alpha \in \Q$:
\begin{equation}
F(S)=\delta
\end{equation}
where $\delta$ is the Dirac distribution. That follows in fact from the ergodic Birkhoff theorem, but the speed of the
convergence to zero of $\frac{S_n(x,y;\alpha)}{n}$ is very slow. 

\section[The irrational case]{The case of irrational $\beta $ 
in terms of Gaussian sums}
\setcounter{equation}{0}

In this section we shall proceed with a different type of estimate of 
$E(S_n^{2q}(\alpha))$ for $\alpha \in  \R^{\ast}$. In particular, we emphasize the
difference between the rational case of section
2 and the irrational  case $\alpha \in \R \setminus \Q$.
For this purpose, we shall establish a generalization of the duality formula for
one-dimensional Gaussian sums, given in the celebrated work of
Hardy and Littlewood [HL]. This
problem consists in estimating the behavior of the many variable
sums, for large $n$:
\begin{equation}
\sum_{k_1=0}^{n-1} \ldots \sum_{k_{2q}=0}^{n-1} e^{i\frac{\alpha}{4}
\sum_{i=2}^{2q}\sum_{j=2}^{2q}
\frac{P_iP_j}{\sum_{i=2}^{2q}P_i}(k_i-k_j)^2}
\end{equation}
where $P_i\in \Zz$, $i=2,\ldots,2q$ are parameters (see Eq.\ref{eq3207}). The above sum bears 
over integer in  indices of $\N^{2q}$ formed by intersection of the hypercube 
 of total volume $n^{2q}$ and the hyperplane  defined by the equation:
\begin{equation}
\sum_{i=2}^{2q}P_i(k_i-k_1)=0
\end{equation}
This condition can be recast as an integral
using the formulae:

\begin{eqnarray}
\int_0^1e^{2\pi i xP}dx &=&1 \,\, if \,\, P=0 \nonumber \\
&=& 0\,\, if \,\, P\in \N^{\ast}
\end{eqnarray}
Define now,  the quantity:
\begin{eqnarray}
\label{eq3304}
\lefteqn{\xi_n[x,\beta;P_2,\ldots, P_{2q}]= }\\
& & \sum_{k_1=0}^{n-1}\ldots \sum_{k_{2q}=0}^{n-1}e^{i\pi \beta
\sum_{i=2}^{2q} \sum_{j=2}^{2q} \frac{P_iP_j}{
\sum_{i=2}^{2q}P_i}(k_i-k_j)^2} e^{2i\pi x
\sum_{i=2}^{2q}P_i(k_i-k_1)} \nonumber
\end{eqnarray}
Then, the $2q$\textsuperscript{th} (see Eq.\ref{eq3207})  moment may be
rewritten as:

\begin{eqnarray}
\label{eq3305}
\lefteqn{E[S_n^{2q}(\frac{\pi \beta}{4})]=  (-1)^q\left( \frac{4}{\pi}\right)^q  }\\
& &\sum_{P_i \in \Zz} \frac{1}{\prod_{i=2}^{2q}P_i(\sum_{i=2}^{2q}P_i)}
\int _0^1\xi_n [x,\beta;P_2,\ldots, P_{2q}]dx \nonumber
\end{eqnarray}

As $ \xi_n[x,\beta;P_2,\ldots, P_{2q}] $ are analogous to $2q$ variables theta functions [M], we shall express
the formula Eq. \ref{eq3305} in terms of Gaussian sums $
^d\sigma_{A_{\vec{n}}}^{[\vec{a},\vec{b}]} (\Omega,\vec{\theta})$.
\begin{definition}
we  define the d-dimensional Gaussian sums:
\begin{eqnarray}
 ^d\sigma_{A_{\vec{n}}}^{[\vec{a},\vec{b}]} (\Omega,\vec{\theta}) =\sum_{\vec{k} \in A_{\vec{n}}}  e^{ i\pi \, 
^t(\vec{k}-\frac{\vec{a}}{2})\Omega
(\vec{k}-\frac{\vec{a}}{2}) }e^{2i\pi
\,^t(\vec{k}-\frac{\vec{a}}{2})(\vec{\theta}-\frac{\vec{b}}{2})}
\end{eqnarray}
Where $\Omega$ is a $d$ square matrix with real coefficients: $\Omega
\in M_d(\R)$, $\vec{\theta} \in Vect_d(\R)$ such that $\forall
i=1,\ldots,d$, the  component $\theta_i$ of $\vec{ \theta}$ is
restricted to $0<\theta_i<1$. Here we denote by $^t\vec{v}$ the transpose of a vector $\vec{v}$. Moreover $\vec{a}$ and
$\vec{b}$ are elements of $Vect_d(\{0,1\})$. $A_{\vec{n}}$, with
$\vec{n}=(n_1,\ldots,n_d)$,  is the hyper-rectangle with integer
sites in $\N^d$ defined by:

Any $ \vec{k} \in A_{\vec{n}}$ is such that  $\forall
i=1,\ldots,d$, $0\leq k_i\leq n_i-1$, $k_i \in \N$.
\\ With this
notation:
\begin{eqnarray}
\label{eq3304}
 \xi_n[x,\beta;P_2,\ldots, P_{2q}]=
\sum_{k_1=0}^{n-1}e^{-2i\pi x (\sum_{i=2}^{2q}P_i) k_1}.
{^{2q-1}}\sigma_{A_{\vec{n}}}^{[\vec{0},\vec{0}]}
(\Omega,\vec{\theta})
\end{eqnarray}

where:
\begin{itemize}
\item
$A_{\vec{n}}$ defined by the $2q-1$-vector $\vec{n}$:
$\vec{n}=(n-1,\ldots,n-1)$
\item the $2q-1$-square matrix $\Omega$:

\begin{eqnarray}
\Omega= \frac{\beta}{\sum P_i} \left[
\begin{array}{cccc}
P_2(\sum_{i=2}^{2q}P_i-P_2) & -P_2P_3 &\ldots &-P_2P_{2q}\\
-P_2P_3 &P_3(\sum_{i=2}^{2q}P_i-P_3) &\vdots &-P_3P_{2q}\\
\vdots &\vdots &\ddots &\vdots \\
-P_2P_{2q} & -P_3P_{2q} &\ldots &P_{2q}(\sum_{i=2}^{2q}P_i-P_{2q})
\end{array}
\right]\nonumber \\
\end{eqnarray}
\item the $2q-1$-vector $\theta$:
\begin{eqnarray}
\vec{\theta}=x \left[
\begin{array}{c}
P_2\\
P_3\\
\vdots\\
P_{2q}
\end{array}
\right] \in Vect_{2q-1}(\R)
\end{eqnarray}

\item
$\vec{a}=\vec{b}=\vec{0}$

\end{itemize}

\end{definition}
In order to compute the above sum, we shall first establish the duality formula:
$\\$
\begin{theor}
[duality formula]
\label{theo3}
If $\Omega\in M_d(]0,1[)_{sym,inv}$ ( the d-dimensional symmetric and invertible
matrices with values in ]0,1[),  $\vec{\theta} \in
Vect_d(]0,1[)$, and
$(\vec{a},\vec{b}) \in Vect_d(\{0,1\})$, then\,
\begin{eqnarray}
\lefteqn{ ^d\sigma_{\mathcal{D}}^{[\vec{a},\vec{b}]} (\Omega,\vec{\theta}) =}\\
& & \left( \frac{i}{\sqrt{\det{\Omega}}}\right)^\frac{d}{2} e^{-i\pi\,
^t\vec{\theta} \Omega^{-1} \vec{\theta}   } \left( \sum_{0\leq i_1,
\ldots, i_d \leq d}  \mathcal{O}(1)^{i_1+\ldots+i_d}
.{^d}\sigma_{\mathcal{P}_{i_1}\ldots \mathcal{P}_{i_d}(\Omega
\mathcal{D})}^{[\vec{b},\vec{a}]}
(-\Omega^{-1},\Omega^{-1}\vec{\theta})    \right) \nonumber
\end{eqnarray}
\end{theor}
where $\mathcal{D}$ is a domain of $\N^d$ and $\Omega(\mathcal{D})$
is the set of integers sites of $\N^d$ contained in the image of
$\mathcal{D}$ by $\Omega$ and $\mathcal{P}_i$ is the projection on
$\{k_i=0\}$. Moreover we take the convention $\mathcal{O}(1)^0=1$.

This duality formula will be extended to $\Omega \in
M_d(\R)_{sym,inv}$ (d-dimensional real symmetric and invertible matrices) and later we shall treat
the case of singular
$\Omega$, i.e.: $\det(\Omega)=0$.

\subsection{Results of Hardy and Littlewood on  one-dimensional Gaussian sums}
\setcounter{equation}{0}

Following [HL], let us introduce:
\begin{eqnarray}
C_n^2(x,\theta) & = &\sum_{k=0}^{n-1}e^{i\pi x(k-1/2)^2}\cos(2k-1)\pi \theta \nonumber\\
C_n^3(x,\theta) & = &\sum_{k=0}^{n-1}e^{i\pi xk^2}\cos 2k\pi \theta \nonumber\\
C_n^4(x,\theta) & = &\sum_{k=0}^{n-1}(-1)^ke^{i\pi x k^2}\cos 2k\pi
\theta
\end{eqnarray}
Then, the duality formula for the Gaussian sum $C_n^3(x,\theta)$ (see Eq. \ref{eq3317} below ) can
be obtained by application of the residue theorem in the formula [HL]:
\begin{eqnarray}
C_n^3(x,\theta)=\sqrt{\frac{i}{x}}e^{-i\pi \frac{\theta^2}{x}}
C_{nx}^3(-\frac{1}{x},\frac{\theta}{x}) +i\int_0^\infty
\frac{e^{-\pi x (n^2-t^2)}}{\sinh\pi t}Q_n(\theta,x,t) dt
\end{eqnarray}
with $0<x<1$ and $0<\theta<1$. The function $Q_n(\theta, x,t)$ is
given by:
\begin{eqnarray}
Q_n(\theta,x,n) & = & \cos(2\pi n \theta) \cosh(2\pi \theta t) \sinh((2nx-2k-1)\pi t) \\
&+ &  i \sin( 2\pi n\theta) \sinh(2\pi \theta t)
\cosh((2nx-2k-1)\pi t) \nonumber
\end{eqnarray}
Observe that the integral in this equation may be reduced to
integrals of the type :
\begin{eqnarray}
\int_0^\infty e^{i\pi x (n^2-t^2)} e^{2\pi \theta
t}\frac{\sinh(\alpha \pi t)}{\sinh(\pi t)}dt.
\end{eqnarray}

From the theorem on intermediate values we have: $\exists T$
bounded, such that $\forall \theta \in ]0,1[$, the previous
integral is
\begin{eqnarray}
\mathcal{O} (1) e^{-\pi n^2 x} \int_0^{T(\theta)}  e^{i\pi x t^2}
e^{2\pi \theta t}dt
\end{eqnarray}
where $\mathcal{O}(1)$ means a bounded function with respect to
$\theta$. This integral depends on $\theta$ but remains bounded.
The validity of this theorem is guaranteed since
$\frac{\sinh(\alpha \pi t)}{\sinh(\pi t)}$ is a positive
decreasing function of $\theta$ on $\R^{+}$. Its value is:
\begin{eqnarray}
\mathcal{O} (1) \sqrt{\frac{i}{x}}e^{-\pi \frac{\theta^2}{
x}}\left( \mbox{erf}(T\sqrt{i\alpha}
-\frac{b}{\sqrt{i\alpha}})-\mbox{erf}(\frac{b}{\sqrt{i\alpha}}) \right)
\end{eqnarray}
where the function $\mbox{erf}$ (the error function) is defined as
$\mbox{erf}(z) \mapsto \int_0^z e ^{-i\pi v^2}dv$ This leads to a
more precise form of the result of Hardy et Littelwood. A similar
proof can be performed for $C_n^2$ and $C_n^4$. In fact we have obtained
the proposition:

\begin{proposition}
\label{prop1_b} We have the following duality formulas: $\forall
0<x<1$ and  $\forall 0<\theta<1$
\begin{eqnarray}
\label{eq3317}
C_n^2(x,\theta) & = &\sqrt{\frac{i}{x}}e^{-i\pi \frac{\theta^2}{x}} \left( C_{nx}^4 \left(-\frac{1}{x},\frac{\theta}{x} \right) +\mathcal{O}(1)\right) \nonumber\\
C_n^3(x,\theta) & = &\sqrt{\frac{i}{x}}e^{-i\pi \frac{\theta^2}{x}} \left( C_{nx}^3 \left(-\frac{1}{x},\frac{\theta}{x} \right) +\mathcal{O}(1)\right) \nonumber\\
C_n^4(x,\theta)& =
&\sqrt{\frac{i}{x}}e^{-i\pi\frac{\theta^2}{x}}\left( C_{nx}^2
\left(-\frac{1}{x},\frac{\theta}{x} \right) +\mathcal{O}(1)\right)
\end{eqnarray}
\end{proposition}In their work, Hardy and Littlewood, have stated that analogous sums with sine
instead cosine obey also duality formulas which can be established
in a similar way. However as already seen, the parity of the
cosine function does play an important role. When the cosine is
replaced by the sine, there appears Fresnel functions which need
additional treatment. We offer here an alternative proof. Define:
\begin{eqnarray}
C_{q,q+r}^3(x,\theta)=\sum_{k=q+1}^{q+r} e^{i\pi x k^2} \cos(2k\pi
\theta) = C_{q+r}^3(x,\theta)-C_{q}^3(x,\theta)
\end{eqnarray}
Applying the duality formula on $C_{q}^3(x,\theta)$ we find:
\begin{eqnarray}
C_{q,q+r}^3(x,\theta)  = \sqrt{\frac{i}{x}}e^{-i\pi
\frac{\theta^2}{x}} \left( C_{nq,n(q+r)}^3
\left(-\frac{1}{x},\frac{\theta}{x} \right)
+\mathcal{O}(1)\right).
\end{eqnarray}
But we also have:
\begin{eqnarray}
C_{q,q+r}^3(x,\theta)  = \sum_{k=q+1}^{q+r} e^{i\pi x
k^2}\cos(2\pi k\theta)= \sum_{k=1}^{q+r} e^{i\pi x
(k+q)^2}\cos2\pi(k+q)\theta \nonumber\\
\end{eqnarray}
after expansion of the cosine, we obtain:
\begin{eqnarray}
\lefteqn{C_{q,q+r}^3(x,\theta) = \frac{e^{-\pi x q^2}}{2}\sum_{k=1}^{r} e^{i\pi x k^2}( \cos{2\pi\theta q}} \\
&  &[\cos{2\pi(xq+\theta)k} +\cos{2\pi(xq-\theta)k}+i (\sin{2\pi(xq+\theta)k}  +\sin{2\pi(xq+\theta)k})]  \nonumber\\
& +&  \sin{2\pi\theta q} \nonumber\\
& & [\sin{2\pi(xq+\theta)k} -\sin{2\pi(xq-\theta)k}-i (\sin{2\pi(xq+\theta)k} + \sin{2\pi(xq+\theta)k})])\nonumber
\end{eqnarray}

Assume now $x\in \Q$, such that $x=a/b$, where $a$ and $b$ are
prime numbers. We choose now $e_1$ as multiple of $b$ and let
$e(xq)$ label the integer part of $xq$:
\begin{eqnarray}
C^3_{q,q+r}(x,\theta) & = &  (-1)^{q e(xq)} \\
& & \left[ \cos 2\pi q \theta \sum_{k=0}^{r-1} e^{i \pi x k^2}\cos
2\pi \theta k -\sin 2\pi q \theta  \sum_{k=0}^{r-1} e^{i \pi x
k^2}\sin 2\pi \theta k  \right] \nonumber.
\end{eqnarray}
Using the duality formula we obtain:
\begin{eqnarray}
C^3_{q,q+r}(x,\theta) &  = & (-1)^{q e(xq)} \cos 2\pi q \theta    \sqrt{\frac{i}{x}} e^{-i\pi \frac{\theta^2}{x}} \left(  C_{xr}^3\left( -\frac{1}{x},\frac{\theta}{x}\right)+\mathcal{O}(1) \right)    \nonumber \\
& & - (-1)^{q e(xq)} \sin 2\pi q \theta  \sum_{k=0}^{r-1} e^{i \pi
s k^2}\sin 2\pi \theta k.
\end{eqnarray}
Moreover we have

\begin{eqnarray}
 & &\sqrt{\frac{i}{x}} e^{-i\pi \frac{\theta^2}{x}} \left(  C_{xr}^3\left( -\frac{1}{x},\frac{\theta}{x}\right)+\mathcal{O}(1) \right) = \sqrt{\frac{i}{x}} e^{-i\pi \frac{\theta^2}{x}}  \sum_{k=0}^{e(xr)-1}  \sqrt{\frac{i}{x}} e^{-i\pi \frac{\theta^2}{x}}  \nonumber\\
& & (-1)^{q e(xq)}e^{-i\pi \frac{(k+e(xq))^2}{x}}  \cos 2\pi \frac{k+e(xq)}{x}    \theta  \left\lbrack \cos 2\pi  \frac{e(xq)}{x}\theta  \sum_{k=0}^{e(xr)-1} e^{-i \pi \frac{k^2}{x}}  \right.\nonumber \\
& &\left.  \cos 2\pi \frac{k}{x} \theta - \sin 2\pi  \frac{e(xq)}{x}\theta  \sum_{k=0}^{e(xr)-1}e^{-i \pi \frac{k^2}{x}} \sin 2\pi \frac{k}{x} \theta    + \mathcal{O}(1)   \right\rbrack
\end{eqnarray}
This leads to the equality:
 \begin{eqnarray}
& &  \cos 2\pi \theta q  \left( C_{xr}^3 \left( \frac{-1}{x}, \frac{\theta}{x}\right)  + \mathcal{O}(1) \right)   -\sqrt{\frac{x}{i}} \sin 2\pi q \theta  \sum_{k=0}^{e(xr)-1} e^{i\pi \frac{\theta^2-k^2}{x}} \sin 2\pi \frac{k}{x} \theta  \nonumber \\
& &=  C_{xr}^3 \left( \frac{-1}{x}, \frac{\theta}{x}\right) \cos 2\pi  \frac{e(xq)}{x}\theta  -  S_{xr}^3 \left( \frac{-1}{x}, \frac{\theta}{x}\right) \sin 2\pi  \frac{e(xq)}{x}\theta + \mathcal{O}(1)  \nonumber \\
\end{eqnarray}
when $q$ is given as a multiple of $b$, we  get simply:

\begin{eqnarray}
S_{r}^3(x,\theta)=  \sqrt{\frac{i}{x}}  e^{-i\pi \frac{\theta^2}{x}} \left[ S_{xr}^3 \left( \frac{-1}{x}, \frac{\theta}{x}\right)+\mathcal{O}(1) \right]
\end{eqnarray}
$ \forall \theta$ such that $2\pi q \theta \not= \pi k$ or $\theta \not\equiv \frac{k}{2q} \mod(1)$. Let us now call
\begin{eqnarray}
\zeta_q= \left\lbrace \theta \in ]0,1[ \quad  \exists k \in \N \quad  \mbox{such that} \quad  \theta \equiv \frac{k}{2q} \mod(1) \right\rbrace
\end{eqnarray}
But $\forall \theta$, $\exists q^{\prime}$ a multiple of $b$ such that $\theta \not \in \zeta_{q^{\prime}}$.

Hence, $\forall \theta \in ]0,1[$ and $\forall x\in \Q$, we obtain
the same duality formula . Finally $S_r(x,\theta)$ is a continuous
function and bicontinuous since it is a finite sum of bicontinuous
functions. As $\Q$ is dense $ \R$. $\forall x\in ]0,1[$, $\exists
x_n \in \Q\setminus ]0,1[$, $n\in \N$ such that $\lim_{n\to
\infty}x_n=x$ and by continuity we obtain the limits:
\begin{eqnarray}
\lim_{n\to\infty} S_{r}^3(x_n,\theta) & = & S_{r}^3(x,\theta) \nonumber \\
\lim_{n\to\infty} S_{xr}^3 \left( \frac{-1}{x_n}, \frac{\theta}{x_n}\right) & = & S_{xr}^3 \left( \frac{-1}{x}, \frac{\theta}{x}\right)
\end{eqnarray}
Doing the same proof for $S_n^2(x,\theta)$ and $S_n^4(x,\theta)$ we get the following duality formulas:
\begin{eqnarray}
S_n^2(x,\theta) & = &\sqrt{\frac{i}{x}}e^{-i\pi \frac{\theta^2}{x}} \left( S_{nx}^4 \left(-\frac{1}{x},\frac{\theta}{x} \right) +\mathcal{O}(1)\right) \nonumber\\
S_n^3(x,\theta) & = &\sqrt{\frac{i}{x}}e^{-i\pi \frac{\theta^2}{x}} \left( S_{nx}^3 \left(-\frac{1}{x},\frac{\theta}{x} \right) +\mathcal{O}(1)\right) \nonumber\\
S_n^4(x,\theta) & = &\sqrt{\frac{i}{x}}e^{-i\pi \frac{\theta^2}{x}} \left( S_{nx}^2 \left(-\frac{1}{x},\frac{\theta}{x} \right) +\mathcal{O}(1)\right)
\end{eqnarray}
These results are used to show that, for the following Gaussian sums:
\begin{eqnarray}
^1\sigma_n^{[a,b]}(x,\theta) = \sum_{k=0}^{n-1} e^{-i\pi x (k-a/2)^2} e^{2i\pi (k-a/2)(\theta-b/2)}
\end{eqnarray}
with $a=1,0$ and $b=0,1$, we have $\forall x \in ]0,1[$ and $\theta \in ]0,1[$:
\begin{eqnarray}
^1\sigma_n^{[a,b]}(x,\theta) = \sqrt{\frac{i}{x}}e^{-i\pi \frac{\theta^2}{x}} \left( {^1}\sigma^{[a,b]}_{nx} \left(-\frac{1}{x},\frac{\theta}{x} \right) +\mathcal{O}(1)\right)
\end{eqnarray}

we summurise these results in the following:

\begin{theor}
\label{theo4}
$\forall x\in ]0,1[, \forall \theta \in ]0,1[, \forall a,b=0,1$:
\begin{eqnarray}
^1\sigma_n^{[a,b]}(x,\theta) & = & \sqrt{\frac{i}{x}}e^{-i\pi \frac{\theta^2}{x}} \left( {^1}\sigma^{[b,a]}_{nx} \left(-\frac{1}{x},\frac{\theta}{x} \right) +\mathcal{O}(1)\right) \nonumber\\
C_n^{[a,b]}(x,\theta) & = & \sqrt{\frac{i}{x}}e^{-i\pi \frac{\theta^2}{x}} \left(
C^{[b,a]}_{nx} \left(-\frac{1}{x},\frac{\theta}{x} \right) +\mathcal{O}(1)\right) \nonumber\\
S_n^{[a,b]}(x,\theta) & = & \sqrt{\frac{i}{x}}e^{-i\pi \frac{\theta^2}{x}} \left(
S^{[b,a]}_{nx} \left(-\frac{1}{x},\frac{\theta}{x} \right) +\mathcal{O}(1)\right)
\end{eqnarray}
\end{theor}
Upon application of duality $\forall x\in ]0,1[, \forall \theta
\in ]0,1[, \forall a,b=0,1$, we have:
\begin{eqnarray}
^1\sigma_n^{[a,b]}(-x,\theta) & = & \left[ ^1\sigma_n^{[a,b]}(x,\theta)  \right]^{\ast} \nonumber \\
& = & \left[ \sqrt{\frac{i}{x}}e^{-i\pi \frac{\theta^2}{x}} \left( {^1}\sigma^{[a,b]}_{nx} \left(-\frac{1}{x},\frac{\theta}{x} \right) +\mathcal{O}(1)\right)  \right]^{\ast} \nonumber \\
& = &\frac{e^{-i\frac{\pi}{4}}}{\sqrt{x}} e^{i\pi \frac{\theta^2}{x}}\left[  {^1}\sigma^{[b,a]}_{nx} \left(-\frac{1}{x},\frac{\theta}{x} \right) +\mathcal{O}(1) \right]^{\ast} \nonumber \\
& = &\frac{e^{-i\frac{\pi}{4}}}{\sqrt{-x}} e^{i\pi \frac{\theta^2}{-x}}\left[  {^1}\sigma^{[b,a]}_{-nx} \left(-\frac{1}{-x},\frac{\theta}{-x} \right) +\mathcal{O}(1) \right]^{\ast}
\end{eqnarray}
we extend the validity of the duality formula to $x\in \R \setminus \N$.
Using the convention $\sqrt{-1} =i$, the identity formula can be
extended for $-1<x<1$ and $x\not=0$. All this remain valid for
$C_n$ and $S_n$, since we have the identity:
\begin{eqnarray}
^1\sigma_n^{[a,b]}(x,\theta)= C_n^{[a,b]}(x,\theta)+i.S_n^{[a,b]}(x,\theta)
\end{eqnarray}
Finally we also have the formula:
\begin{eqnarray}
^1\sigma_n^{[a,b]}(x+1,\theta)=\sqrt{i}^a {^1}\sigma_n^{[a,b^{\prime}]}(x,\theta)
\end{eqnarray}
with $b^{\prime}\equiv b+a \mod(1)$, it may be used to extend the
validity domain of the formula into $x\in \R \setminus \Z$. If
$x\in \Z$, the series reduces to a geometric series. The same
results hold  for $C_n$ and $S_n$.

\begin{remark}
\label{rem2}
\end{remark}
If $x\in \Q$, there exists a discrete duality formula (Shaar's
formula). This relation is actually the same as our formulas
without the term $\mathcal{O}(1)$. The formulas are used in
successive iterations until the summations are limited to a small
number. The domain of summation is reduced at each step. However
the result depends on its expansion as a continuous fraction (see
article Hardy and Littlewood). This last problem is closely
related to the Gauss transform $T(x) \equiv \frac{1}{x} \mod(1)$
for which there exists infinite generating partitions. So for
example if $x$ is irrational, but admits a decomposition into
continuous bounded fractions we have:
\begin{eqnarray}
^1\sigma_n^{[a,b]}(x,\theta)\simeq \mathcal{O}(\sqrt{n})
\end{eqnarray}
a homogenous function in $\theta$. But if $x \in \Q$:
\begin{eqnarray}
^1\sigma_n^{[a,b]}(x,\theta)\simeq \mathcal{O}(n).
\end{eqnarray}

\section{Two-dimensional gaussians sums}
\subsection{Two-dimensional Gaussian sums with non-degenerate matrix $\Omega$}
\setcounter{equation}{0}

Once this case treated, one can extend the results to
Gaussian sums of arbitrary dimensions, provided some delicate
points are worked. Consider the Gaussian sum in two dimensions:
\begin{eqnarray}
\label{2eq4b}
^2\sigma_{A_{\vec{q}}}^{[0,0]}(\Omega,\vec{\theta})
=\sum_{k_1=0}^{q_1-1} \sum_{k_2=0}^{q_2-1} e^{i\pi
(\omega_1k_1^2+\omega_2k_2^2+2\omega k_1k_2)}e^{2i \pi
(k_1\theta_1+k_2\theta_2)}
\end{eqnarray}
where $\vec{q}=(q_1,q_2)$ and $\Omega \in M(]0,1[)_{sym}$ with
\begin{eqnarray}
\Omega =
\left[ \begin{array}{cc}
\omega_1 &\omega \\
\omega &\omega_2
\end{array} \right]
\end{eqnarray}

$A_{\vec{q}}$ labels integer sites of rectangles of dimensions
$q_1\times q_2$. We first assume that
$0<\omega q_1 +\theta_1<1$ and $0<\omega q_2+\theta_2<1$. As $q_1$
and $q_2$ are to be large numbers, one may call these matrices
{\it weak coupling} matrices. If $\omega=0$, this Gaussian sum
goes back to a product of one-dimensional Gaussian sums. In
general:
\begin{eqnarray}
^2\sigma_{A_{\vec{q}}}^{[0,0]}(\Omega,\vec{\theta}) =
\sum_{k_1=0}^{q_1-1} e^{i\pi (\omega_1k_1^2 +2\theta_1 k_1)}
\sum_{k_2=0}^{q_2-1} e^{i\pi (\omega_2k_2^2+2\omega k_1k_2+2
\theta_2 k_2)}
\end{eqnarray}
We shall establish the duality formula successively with respect to ($k_1$, $k_2$).
If the duality transformation is performed on the $k_2$ sum one gets:
\begin{eqnarray}
^2\sigma_{A_{\vec{q}}}^{[0,0]}(\Omega,\vec{\theta}) & = &\sum_{k_1=0}^{q_1-1} e^{i\pi (w_1k_1^2 +2\theta_1 k_1)}
e^{-i\pi \frac{(\omega k_1+ \theta_2)^2}{\omega_2}}  \nonumber \\
&& \sqrt{\frac{i}{\omega_2}}  \left( \sum_{k_2=0}^{\omega_2 (q_2-1)}
e^{-i\pi \frac{k_2^2}{\omega_2}} e^{2i\pi \frac{(\omega k_1+\theta_2)^2}{\omega_2}}  + \mathcal{O}(1) \right) \nonumber \\
& = &  \sqrt{\frac{i}{\omega_2}}  e^{-i\pi \frac{ \theta_2^2}{\omega_2}} \left( \sum_{k_2=0}^{\omega_2 (q_2-1)} e^{-i\pi  \frac{ k_2^2}{\omega_2}}
e^{2i\pi  \frac{ \theta_2}{\omega_2}k_2} e^{2i\pi  ( \theta_1-\theta_2\frac{\omega}{\omega_2}+\frac{\omega}{\omega_2}k_2)k_1} \right. \nonumber\\ & &
\left.\sum_{k_1=0}^{q_1-1} e^{i\pi \frac{\omega_1 \omega_2 -\omega^2}{ \omega_2}k_1^2}  \right) \nonumber \\ & + & \mathcal{O}(1) 
\sqrt{\frac{i}{\omega_2}}  e^{-i\pi \frac{ \theta_2^2}{\omega_2}}  \sum_{k_1=0}^{q_1-1}  e^{i\pi \frac{\omega_1 \omega_2 -\omega^2}{ \omega_2}k_1^2}
e^{2i\pi (\theta_1-\frac{\omega}{\omega_2}\theta_2) k_1}
\end{eqnarray}
when $\Omega$ is invertible matrix ($\det(\Omega)\not=0$), we can apply
further the duality transformation on the $k_1$ sum and obtain:
\begin{eqnarray}
^2\sigma_{A_{\vec{q}}}^{[0,0]}(\Omega,\vec{\theta}) & = &  \sqrt{\frac{i}{\omega_2}} e^{-i\pi \frac{ \theta_2^2}{\omega_2}} \left(  \sum_{k_2=0}^{\omega_2q_2-1}  e^{-i\pi \left( \frac{k_2^2}{\omega_2}+2\frac{ \theta_2^2}{\omega_2}k_2 \right) }
e^{-i\pi \frac{\omega_2\theta_2-\omega \theta_2+\omega k_2}{\omega_2(\omega_1\omega_2-\omega^2)}k_1}\right)  \nonumber  \\
&\times & \left( \sum_{k_1=0}^{\left( \omega_1- \frac{\omega^2}{\omega_2} \right) q_1-1}  e^{-i\pi \frac{\omega_2}{\omega_1\omega_2-\omega^2}k_1^2}  e^{2i\pi \frac{\omega_2\theta_1-\omega \theta_2+\omega k_2}{\omega_2(\omega_1\omega_2-\omega^2)}k_1}\right)  \nonumber  \\
& + &  \mathcal{O}(1)\sqrt{\frac{i}{\omega_2}} e^{-i\pi \frac{ \theta_2^2}{\omega_2}} e^{-i\pi \frac{\omega_2}{\omega_1\omega_2-\omega^2}(\theta_2-\frac{\omega}{\omega_2}\theta_1)} \\
& \times & \left( \sum_{k_2=0}^{\left( \omega_1- \frac{\omega^2}{\omega_2} \right) q_2-1}  e^{-i\pi \frac{\omega_2}{\omega_1\omega_2-\omega^2}k_1^2}  e^{2i\pi \frac{\omega_2\theta_2-\omega \theta_2}{\omega_2(\omega_1\omega_2-\omega^2)}k_1} +\mathcal{O}(1) \right)  \nonumber
\end{eqnarray}

Now taking into account the assumptions on $\omega$, the summation
over the variable $k_1$ from 0 to $\left( \omega_1-
\frac{\omega^2}{\omega_2} \right) q_1-1$  is the same as the
summation on  $k_1=0,\ldots, \omega_1q_1-1$. This fact occurs also
for the $k_2$ such that the above expression becomes:

\begin{eqnarray}
&  & ^2\sigma_{A_{(q_1,q_2)}}^{[0,0]}(\Omega,\vec{\theta})  = \frac{ie^{-i\pi\,
^t\vec{\theta} \Omega^{-1} \vec{\theta}   }}{\sqrt{det(\Omega)}} \left\lbrack {^2}\sigma_{A_{( \omega_1 q_1, \omega_2q_2)}}^{[0,0]}(-\Omega^{-1},\Omega^{-1}\vec{\theta})   + \right. \\
& & \left. \mathcal{O}(1)\left( {^2}\sigma_{\mathcal{P}_1[A_{(\omega_1 q_1,\omega_2q_2)}]}^{[0,0]}(-\Omega^{-1},\Omega^{-1}\vec{\theta})+{^2}\sigma_{\mathcal{P}_2[A_{(\omega_1 q_1,\omega_2q_2)}]}^{[0,0]}(-\Omega^{-1},\Omega^{-1}\vec{\theta}) +1 \right) \right\rbrack \nonumber
\end{eqnarray}
where $\mathcal{P}_1$, (resp. $\mathcal{P}_2$) is the projection
onto the plane $\{k_1=0\}$ ( resp. $\{k_2=0\}$). We notice also
that the duality transformation replaces essentially sums over $k$
by other sums over $k$ with an error of order $1$. For example
$\sum_{k_1,k_2}$ is replaced by $(\sum_{k_1}+\mathcal{O}(1))$
$(\sum_{k_2}+\mathcal{O}(1)) \simeq \sum_{k_1,k_2}+
\mathcal{O}(1)\sum_{k_1}+\mathcal{O}(1)\sum_{k_2}
+\mathcal{O}^2(1)$. Moreover, there is a simplification since
\begin{eqnarray}
^2\sigma^{[0,0]}_{\mathcal{P}_1[A(\omega q_1,\omega_2 q_2)]}(-\Omega^{-1},\Omega^{-1} \vec{\theta}) = ^1\sigma^{[0]}_{A(\omega_2 q_2)}\left(-\frac{\omega_1}{\omega_1\omega_2-\omega^2},\frac{\omega_1\theta_2-\omega \theta_1}{\omega_1\omega_2-\omega^2} \right)\nonumber\\
\end{eqnarray}
For sufficiently small $\omega$, integer sites such that $0\leq
k_1\leq \omega_1q_1$ and $0\leq k_2\leq \omega_2q_2$
 are those with $\vec{k}=(k_1,k_2) \in \Omega( A_{\vec{q}})$. Thus:
\begin{eqnarray}
^2\sigma_{A_{\vec{q}}}^{[0,0]}(\Omega,\vec{\theta}) & = & \frac{ i e^{-i\pi\,
^t\vec{\theta} \Omega^{-1} \vec{\theta}   }}{\sqrt{det(\Omega)}} \left\lbrack {^2}\sigma_{\Omega
A_{\vec{q}}}^{[0,0]}(-\Omega^{-1},\Omega^{-1}\vec{\theta}) +
\mathcal{O}(1).{^2}\sigma_{\mathcal{P}_1[\Omega
A_{\vec{q}}]}^{[0,0]}(-\Omega^{-1},\Omega^{-1}\vec{\theta}) \right. \nonumber \\ & + &\left.
\mathcal{O}(1).{^2}\sigma_{\mathcal{P}_2[\Omega
A_{\vec{q}}]}^{[0,0]}(-\Omega^{-1},\Omega^{-1}\vec{\theta}) + \mathcal{O}(1) \right\rbrack
\label{2eq5}
\end{eqnarray}
Now for $\vec{a}$ and $\vec{b}$ non zero (i.e. $a_i=\{0,1\};
b_i=\{0,1\}$, $i=1,2$), we have in a similar manner the lemma
\begin{lemma}
\label{lem5}
$\forall \Omega \in M_2(\R)_{sym}$ an invertible matrix and with
$\omega$ small enough:
\begin{eqnarray}
^2\sigma_{A_{\vec{q}}}^{[\vec{a},\vec{b}]}(\Omega,\vec{\theta})  & = & \frac{ie^{-i\pi\,
^t\vec{\theta} \Omega^{-1} \vec{\theta}   }}{\sqrt{det(\Omega)}} [ {^2}\sigma_{\Omega A_{\vec{q}}}^{[\vec{b},\vec{a}]}(-\Omega^{-1}\Omega^{-1},\vec{\theta})
+ \mathcal{O}(1).{^2}\sigma_{\mathcal{P}_1[\Omega A_{\vec{q}}]}^{[\vec{b},\vec{a}]}(-\Omega^{-1}\Omega^{-1},\vec{\theta}) \nonumber \\
& + & \mathcal{O}(1).{^2}\sigma_{\mathcal{P}_2[\Omega
A_{\vec{q}}]}^{[\vec{b},\vec{a}]}
(-\Omega^{-1}\Omega^{-1},\vec{\theta}) + \mathcal{O}(1)].
\end{eqnarray}

where $^2\sigma_{A_{\vec{q}}}^{[\vec{a},\vec{b}]}(\Omega,\vec{\theta}) =
\sum_{\vec{k}\in A_{\vec{q}}} e^{i\pi {^t}(\vec{k}-\frac{\vec{a}}{2})
\Omega (\vec{k}-\frac{\vec{a}}{2}) } e^{2\pi i {^t}(\vec{\theta}-\frac{\vec{b}}{2})
(\vec{k}-\frac{\vec{a}}{2})}$.
\end{lemma}

\subsection{Two-dimensional Gaussian sums with degenerate matrix $\Omega$}
\setcounter{equation}{0}

In the expression of
$^2\sigma_{A_{\vec{q}}}^{[0,0]}(\Omega,\vec{\theta})$ see Eq
\ref{2eq4b} we perform the duality transformation on $k_2$
sums. This operation yields:

\begin{eqnarray}
 ^2\sigma_{A_{\vec{q}}}^{[0,0]}(\Omega,\vec{\theta}))&=& \sqrt{\frac{i}{\omega_2}} 
e^{-i\pi \frac{\theta_2^2}{\Omega_2}} e^{i\pi (q-1)(\theta_1 -\frac{\omega}{\omega_2}\theta_2)}
 \left( \sum_{k_2=0}^{\omega_2q_2-1}e^{-i\pi \frac{k_2^2}{\omega_2}}
e^{2i\pi (\frac{\theta_2}{\omega_2}+\frac{(q-1)\omega}{2\omega_2})k_2} \right. \nonumber \\
&&  \frac{\sin \pi q_1\left(\theta_1-\frac{\omega}{\omega_2} (k_2-1)\right) }
{\sin \pi \left(\theta_1-\frac{\omega}{\omega_2} (k_2-1)\right)}
+ \left. \mathcal{O}(1) \frac{\sin \pi q_1\left(\theta_1-\frac{\omega}{\omega_2} \right) }{\sin \pi 
\left(\theta_1-\frac{\omega}{\omega_2} \right)}   \right) \nonumber \\
\end{eqnarray}

We observe that for $\Omega$ non invertible, the general form of
the sum

\noindent $e^{i\pi \, quadratic \,\, form}.e^{i\pi \, linear \,\,
form}$ is not preserved under this duality transformation.
Moreover the symmetry $k_1\leftrightarrow k_2$ seems to be
apparently broken since its expression depends on which variable
to be transformed first by duality. This type of formula will be
used later on, although with an appropriate form. In the present
case, it is sufficient to perform successive dualities uniquely on
the variable $k_2$ until the summation on $k_2$ is equal to
$\mathcal{O}(1)$. Then perform at last the geometric summation
over $k_1$. We shall now remove the restriction to small $\omega$ and consider the general case. Let us give the following
definition:
\begin{definition}
An invertible symmetric matrix is called non-diagonal integer,
when there are no integer entries outside its diagonal. This set
of such matrices is denoted by $M_{\Z}^{\ast}(\R)_{inv,sym}$
\end{definition}
 Let
\begin{eqnarray}
f_{[\Omega,\vec{\theta}]}(\vec{k}+\vec{n})=e^{i\pi
{^t}(\vec{k}+\vec{n})\Omega(\vec{k}+\vec{n})}e^{2 i\pi
{^t}(\vec{k}+\vec{n})\vec{\theta}}
\end{eqnarray}
with $\vec{\theta^{\prime}}\equiv  \vec{\theta}+\Omega \vec{n}
\mod 1$, we have
\begin{eqnarray}
f_{[\Omega,\vec{\theta}]}(\vec{k}+\vec{n}) & = & e^{2\pi i {^t}\vec{k}\Omega \vec{n}}f_{[\Omega,\vec{\theta}]}
(\vec{k})f_{[\Omega,\vec{\theta}]}(\vec{n}) \nonumber \\
& = &
f_{[\Omega,\vec{\theta}]}(\vec{n})f_{[\Omega,\vec{\theta^{\prime}}]}(\vec{k})
\end{eqnarray}
So for $\theta_1$ and $\theta_2$ given, let
$(q_1^{\prime},q_2^{\prime})$ such that the duality formula is
valid, we cut the domain $\{0\leq k_1<q_1,\, 0\leq k_2<q_2\}$ into small
rectangles of size $q_1^{\prime} \times q_2^{\prime}$. Let us
consider the set:
\begin{eqnarray}
B(m_1,m_2)& =&\left\lbrace (k_1,k_2)  \in \N^2 \quad \vert \quad  m_1q_1^{\prime} \leq k_1<(m_1+1)q_1^{\prime} \quad \mbox{and} \right.\nonumber\\
&& \left.\quad m_2q_2^{\prime} \leq k_2<(m_2+1)q_2^{\prime} \right\rbrace
\end{eqnarray}
Necessarily $m_1\leq  e(q_1/q_1^{\prime})$ and $m_2\leq
e(q_2/q_2^{\prime})$; here $e(q/q')$ represents the
integer part of $q/q'$. It is sufficient to verify that the
duality formula is proved by pasting together the $B(m_1,m_2)$.
Call $\vec{m}=(m_1,m_2)$. Then
\begin{eqnarray}
\sum_{\vec{k}\in B(\vec{m})} f_{ [\Omega,\vec{\theta}]}(\vec{k}) & = & \sum_{\vec{k}\in B(\vec{0})} f_{[\Omega,\vec{\theta}]}(\vec{k}+\vec{m})]\nonumber\\
& = & f_{[\Omega,\vec{\theta}]}(\vec{m}) \sum_{\vec{k}\in
\Omega B(\vec{0})} f_{[\Omega,\vec{\theta}^"]}(\vec{k})
\end{eqnarray}
with $\vec{\theta}^"=\vec{ \theta}+\Omega\vec{m}-\vec{N}$, where
$\vec{N}$ is a vector with integer coordinates such that $\forall
i $, $\theta^"_i=\theta_i+(\Omega\vec{m})_i-N_i \in ]0,1[$. Now we
perform the duality transformation :
\begin{eqnarray}
& & \sum_{\vec{k}\in B(\vec{m})} f_{[\Omega,\vec{\theta}]}(\vec{k}) =
f_{[\Omega,\vec{\theta}]}(\vec{m})i\frac{e^{-i\pi{^t}\vec{\theta}^{\prime}\Omega^{-1}\vec{\theta}^{\prime}}}
{\det(\Omega)}  \left\lbrack   \sum_{\vec{k}\in \Omega B(\vec{0})} f_{[-\Omega^{-1},\Omega^{-1}\vec{\theta}^"]}
(\vec{k}) \right.\\
& &\left. +\mathcal{O}(1) \left(\sum_{\vec{k}\in \mathcal P_1\Omega B(\vec{0})}
f_{[-\Omega^{-1},\Omega^{-1}\vec{\theta}^"]} (\vec{k})
+\sum_{\vec{k}\in \mathcal P_2\Omega B(\vec{0})}
f_{[-\Omega^{-1},\Omega^{-1}\vec{\theta}^"]}(\vec{k})  +1 \right)\right \rbrack   \nonumber
\end{eqnarray}
with
\begin{eqnarray}
e^{i\pi {^t}\vec{\theta}^{\prime}\Omega \vec{\theta}^{\prime}}=e^{i\pi
{^t}\vec{\theta}\Omega^{-1}\vec{\theta}}f_{[-\Omega^{-1},\Omega^{-1}\vec{\theta}]}(\vec{N})
f_{[-\Omega,-\vec{\theta}]}(\vec{m})
\end{eqnarray}
since $^t\vec{N}\cdot\vec{m} \in \mathbb{N}$ . Finally:
\begin{eqnarray}
 &&\sum_{\vec{k}\in B(\vec{m})} f_{[\Omega,\vec{\theta}]}(\vec{k}) = i
\frac{e^{-i\pi{^t}\vec{\theta}\Omega^{-1}\vec{\theta}}}{\sqrt{\det(\Omega)}}
f_{[-\Omega^{-1},\Omega^{-1}\vec{\theta}]}(\vec{N}) \left\lbrack   \sum_{\overset{\vec{k}\in}{\underset{ \Omega
B(\vec{0})}{}}} f_{[-\Omega^{-1},\Omega^{-1}\vec{\theta}-\Omega^{-1}\vec{N}]}(\vec{k}) \right.  \nonumber\\ &&+
\mathcal{O}(1) \left.\left(\sum_{\overset{\vec{k}\in}{\underset{\mathcal P_1\Omega B(\vec{0})}{}}} f_{[-\Omega^{-1},\Omega^{-1}\vec{\theta}-\Omega^{-1}\vec{N}]}(\vec{k}) +\sum_{\overset{\vec{k}\in}{\underset{ \mathcal P_2\Omega B(\vec{0})}{}}} f_{[-\Omega^{-1},\Omega^{-1}\vec{\theta}-\Omega^{-1}\vec{N}]}(\vec{k})  +1\right) \right\rbrack   \nonumber\\
\end{eqnarray}
or alternatively:
\begin{eqnarray}
& & \sum_{\vec{k}\in B(\vec{m}} f_{[\Omega,\vec{\theta}]}(\vec{k}) =  i \frac{e^{-i\pi{^t}\vec{\theta}\Omega\vec{\theta}}}{\det(\Omega)} f_{[-\Omega^{-1},\Omega^{-1}\vec{\theta}]}(\vec{N})  \left\lbrack  \sum_{\overset{(\vec{k}-\vec{N})\in}{\underset{\Omega B(\vec{0})}{}}} f_{[-\Omega^{-1},\Omega^{-1}\vec{\theta}]}(\vec{N})  \right.\nonumber\\
&& \left.+  \mathcal{O}(1) \left( \sum_{\overset{(\vec{k}-\vec{N})\in}{\underset{ \mathcal P_1\Omega B(\vec{0})}{}}} f_{[-\Omega^{-1},\Omega^{-1}\vec{\theta}}(\vec{k})
+\sum_{\overset{(\vec{k}-\vec{N})\in}{\underset{ \mathcal P_2\Omega B(\vec{0})}{}}} f_{[-\Omega^{-1},\Omega^{-1}\vec{\theta}]}(\vec{k})  +1\right) \right\rbrack   \nonumber\\
\end{eqnarray}
Here $\vec{N}$ is an integer vector such that $\forall i=1,2$,
$|N_i-(\Omega\vec{m})_i|<1$. Thus replacing $\vec{N}$ by
$\Omega\vec{m}$ we make an error in the double sum
$\sum_{k_1,k_2}$ of the order $\mathcal O(1)\sum_{k_1}$ and
$\mathcal O(1)\sum_{k_2}$ and an error in respect to the
$\sum_{k_1}$ ( resp. $\sum_{k_1}$) of order $\mathcal O(1)$. The
index $(\vec{k}- \Omega\vec{m}) \in \Omega B(\vec{0})$ is
equivalent to $\vec{k} \in \Omega(\vec{m}-B(\vec{0}))$ or also
$\vec{k} \in \Omega B(\vec{m}))$ by linearity. For the same
reasons we have, for $i=1,2$, $(\vec{k}-\Omega \vec{m}) \in
\mathcal{P}(\Omega B(\vec{0}))$ is equivalent to $\vec{k} \in
\mathcal{P}(\Omega B(\vec{m}))$ and we have:
\begin{eqnarray}
& &\sum_{k_1=0}^{q_1-1} \sum_{k_2=0}^{q_2-1}  f_{[\Omega,\vec{\theta}]}(\vec{k}) =
\sum_{r=0}^{e(q_1/q^{\prime}_1)}\sum_{s=0}^{e(q_2/q^{\prime}_2)} \sum_{\vec{k}\in B(r,s)} f_{[\Omega,\vec{\theta}]}(\vec{k})
i\frac{e^{-i\pi{^t}\vec{\theta}\Omega\vec{\theta}}}{\sqrt{\det(\Omega)}} \sum_{r=0}^{e(q_1/q^{\prime}_1)} \sum_{s=0}^{e(q_2/q^{\prime}_2)}     \nonumber\\
&&=  \left\lbrack   \sum_{\vec{k} \in \Omega B(r,s) } f_{[-\Omega^{-1},\Omega^{-1}\vec{\theta}]}(\vec{k}) \right.\nonumber\\
&& \left. +\mathcal{O}(1) \left( \sum_{\vec{k}\in \mathcal P_1 \Omega B(r,s)} f_{[-\Omega^{-1},\Omega^{-1}\vec{\theta}]}(\vec{k})
+\sum_{\vec{k}\in \mathcal P_2 \Omega B(r,s)} f_{[-\Omega^{-1},\Omega^{-1}\vec{\theta}]}(\vec{k})  +1\right) \right\rbrack \nonumber\\
&&  =  i \frac{e^{-i\pi{^t}\vec{\theta}\Omega\vec{\theta}}}{\det(\Omega)} \left\lbrack   \sum_{\vec{k} \in \Omega A(q_1,q_2)} f_{[-\Omega^{-1},\Omega^{-1}\vec{\theta}]}(\vec{k}) \right.\\
&&\left.+\mathcal{O}(1) \left( \sum_{\vec{k}\in  \mathcal P_1 \Omega A(q_1,q_2)} f_{[-\Omega^{-1},\Omega^{-1}\vec{\theta}]}(\vec{k})
+ \sum_{\vec{k}\in  \mathcal P_2\Omega A(q_1,q_2)} f_{[-\Omega^{-1},\Omega^{-1}\vec{\theta}]}(\vec{k})  +1 \right) \right \rbrack  \nonumber
\end{eqnarray}
Finally the same calculation shows that for all bounded domains
$\mathcal{D}\in \N^2$, we have the theorem

\begin{theor}[Duality]
\label{theo5}
Let $\Omega \in M^{\ast}_2(]0,1[)_{inv}$, $\mathcal{D}$ the
bounded domain in $\N^2$ we have:
\begin{eqnarray}
& &^2\sigma_{\mathcal{D}}^{[\vec{a},\vec{b}]} (\Omega,\vec{\theta}) =
i \frac{e^{-i\pi{^t}\vec{\theta}\Omega\vec{\theta}}}{\det(\Omega)}
 \left\lbrack  {^2}\sigma_{\Omega \mathcal{D}}^{[\vec{a},\vec{b}]}(-\Omega^{-1},\Omega^{-1}\vec{\theta}) +
 \mathcal{O}(1).{^2}\sigma_{\mathcal P_1(\Omega \mathcal{D})}^{[\vec{a},\vec{b}]}(-\Omega^{-1},
 \Omega^{-1}\vec{\theta}) \right.\nonumber\\
& &\left.+\mathcal{O}(1).{^2}\sigma_{\mathcal P_2( \Omega \mathcal{D})}^{[\vec{a},\vec{b}]}(-\Omega^{-1},\Omega^{-1}\vec{\theta})  +\mathcal{O}(1) \right\rbrack   \nonumber\\
\end{eqnarray}
\end{theor}

This formula can be extended to the set of matrices $\Omega \in
M^{\ast}_2(\R)_{inv}$. This result is given in the next section
where the general case with $d$ dimensions is treated.

\section{$d$-dimensional Gaussian sums: Generalizations}
\setcounter{equation}{0}

In this section, proofs or parts of proofs that are similar to
those in two dimensions will not be repeated. Let us consider the
case where non-diagonal elements $\omega_{ij}$, ($i\not=j$) are
sufficiently small for the same vector $\vec{\theta}$, and where
$\Omega$ is invertible. We shall study the case
$\vec{a}=\vec{b}=\vec{0}$. The general case can be deduced by
analogy. Performing the dual transformation on only one variable $k_m$
with $1\leq m\leq d$ on the sum defining:

\begin{eqnarray}
& & ^d\sigma_{A_{\vec{q}}}^{[\vec{0},\vec{0}]} (\Omega,\vec{\theta}) = \sum_{k_1=0}^{q_1-1}\ldots  \sum_{k_d=0}^{q_d-1} e^{i\pi[\sum_{1\leq i\leq d}\omega_{ii}k_i^2+2\sum_{1\leq i<j\leq d}\omega_{ij}k_ik_j+2\sum_{1\leq i\leq d} k_i\theta_i]} \nonumber \\
& = & \sum_{k_1=0}^{q_1-1}\ldots\sum_{k_m=0}^{\omega_{m,m}q_m-1} \ldots  \sum_{k_d=0}^{q_d-1} e^{i\pi[\sum_{1\leq i\leq d}^{ i\not=m}\omega_{ii}k_i^2+2\sum_{1\leq i<j\leq d}^{i\not=m,\, j\not=m}\omega_{ij}k_ik_j+2\sum_{1\leq i\leq d}^{ i\not=m} k_i\theta_i]}  \nonumber\\
& \times  &  \sum_{k_m=0}^{q_m-1} e^{i\pi
[\omega_{mm}k^2_m+2(\theta_m+\sum_{1\leq i \leq
d}\omega_{mi}k_i)k_m]}
\end{eqnarray}
We get:

\begin{eqnarray}
^d\sigma_{A_{\vec{q}}}^{[\vec{0},\vec{0}]} (\Omega,\vec{\theta}) & = &\sum_{k_1=0}^{q_1-1}\ldots\sum_{\hat{k}_m=0}^{q_m-1} \ldots  \sum_{k_d=0}^{q_d-1} e^{i\pi[\sum_{1\leq i\leq d}^{ i\not=m}\omega_{ii}k_i^2+2\sum_{1\leq i<j\leq d}^{i\not=m,\, j\not=m}\omega_{ij}k_ik_j}  \nonumber\\
&\times & e^{2\sum_{1\leq i\leq d}^{ i\not=m} k_i\theta_i} \sqrt{\frac{i}{\omega_{mm}}}  \sum_{k_i\not=k_m}e^{-\frac{\pi}{\omega_{mm}}(\theta_m+\sum_{1\leq i \leq d}\omega_{mi}k_i)^2} \nonumber\\
& \times & \left(    \sum_{k_m=0}^{\omega_mq_m-1} e^{i\pi
\lbrack-\frac{k^2_m}{\omega_{mm}}+2(\theta_m+\sum_{1\leq i \leq
d}\omega_{mi}k_i)\frac{k_m}{\omega_{mm}}\rbrack } +\mathcal{O}(1) \right) \nonumber\\
\end{eqnarray}
Let us define $D_m$ as the dual transformation operator on the
index $k_m$. We observe that for any function $f$:
\begin{eqnarray}
\mathcal{O}_m(1)f(\ldots,k_m,\ldots)=\mathcal{O}(1)
f(\ldots,k_m=0,\ldots)
\end{eqnarray}
Thus regrouping terms using the previous remark, we have:
\begin{eqnarray}
^d\sigma_{A_{\vec{q}}}^{[\vec{a},\vec{b}]} (\Omega,\vec{\theta}) & = & \sqrt{\frac{i}{\omega_{mm}}}e^{-i\pi\frac{\theta_m^2}{\omega_{mm}}}\sum_{k_1=0}^{q_1-1}\ldots  \left(  \sum_{k_m=0}^{\omega_{mm}q_m-1} +\mathcal{O}_m(1) \right)\ldots \nonumber\\
& &  \sum_{k_d=0}^{q_d-1} e^{-i\pi \left(    \frac{k^2_m}{\omega_{mm}}-\sum_{i\not=m} (\omega_{ii}-\frac{\omega_{mi}^2}{\omega_{mm}})k_i^2 \right)} \\
& \times & e^{i\pi\sum^{i\not=m,\, j\not=m}_{1\leq i<j\leq d}(\omega_{ij}-\frac{\omega_{mi}\omega_{mj}}{\omega_{mm}})k_ik_j}e^{i\pi\sum^{i\not=m}_{1\leq i\leq d}\frac{\omega_{mi}}{\omega_{mm}}k_ik_m}\nonumber\\
& \times & e^{2i\pi\sum^{i\not=m}_{1\leq i\leq
d}(\theta_{1}-\frac{\omega_{mi}}{\omega_{mm}}\theta_m)k_i}e^{2i\pi\sum^{i\not=m}_{1\leq
i\leq d}\frac{\theta_m}{\omega_{mm}}k_m} \nonumber
\end{eqnarray}
Now this relation may be reexpressed using the dualization with respect to the variable $k_m$, $\mathcal{D}_m$, defined by:
\begin{eqnarray}
&& \mathcal{D}_m \left( \sum_{k_1=0}^{q_1-1}\ldots  \sum_{k_d=0}^{q_d-1} e^{i\pi[\sum_{1\leq i\leq d}\omega_{ii}k_i^2+2\sum_{1\leq i<j\leq d}\omega_{ij}k_ik_j+2\sum_{1\leq i\leq d} k_i\theta_i]}  \right) = \nonumber \\
& & \Phi^{\prime} \sum_{k_1=0}^{q_1-1}\ldots\left(  \sum_{k_m=0}^{\omega_{mm}q_m-1}
+\mathcal{O}_m(1) \right)\ldots  \sum_{k_d=0}^{q_d-1}
e^{i\pi[\sum_{1\leq i\leq d}\omega_{ii}^{\prime}k_i^2}\nonumber\\
&\times& e^{+2\sum_{1\leq
i<j\leq d}\omega_{ij}^{\prime}k_ik_j+2\sum_{1\leq i\leq d}
k_i\theta_i^{\prime}]}
\end{eqnarray}
Thus under the action of $\mathcal{D}_m$, matrix elements of
$\Omega$, $\vec{\theta}$ are transformed and a multiplicative
factor is introduced, leading to the following relations :
$\\$
\begin{itemize}
\item $\omega^{\prime}_{ij} = \omega_{ij} -\frac{\omega_{im}\omega_{jm}}{\omega_{mm}}$,  $\forall 1\leq i\leq d$, $\forall 1\leq j\leq d$ with $i\not=m$ and $j \not=m$
\item $\omega^{\prime}_{mm} =  -\frac{1}{\omega_{mm}}$,
\item $\theta^{\prime}_{i}=\theta_i-\frac{\omega_{im}}{\omega_{mm}}\theta_m$ $\forall 1\leq i\leq d$, with $i\not=m$
\item $\theta^{\prime}_{m}=\frac{\theta_{m}}{\omega_{mm}}$
\item $ \Phi^{\prime}=\sqrt{\frac{i}{\omega_{mm}}}e^{-i\pi\frac{\theta_m^2}{\omega_{mm}}}\Phi$ with $\Phi=1$
\end{itemize}
$\\$
and of course we have $\Omega^{\prime}=\mathcal{D}_m(\Omega)$,
$\vec{\theta}^{\prime}=\mathcal{D}_m(\theta)$ and
$\Phi^{\prime}=\mathcal{D}_m(\Phi)$. Under the action of duality
on the index $k_m$ the matrix $\Omega$  becomes a symmetric
matrix. So let us apply this transformation successively as
follows, $1\leq m\leq d$.
\begin{eqnarray}
\left[
\begin{array}{c}
\Omega \\
\vec{\theta} \\
1
\end{array}
\right] = \left[
\begin{array}{c}
\Omega_0 \\
\vec{\theta}_0 \\
\Phi_0
\end{array}
\right] \underset{\mathcal{D}_1}{\longmapsto} \left[
\begin{array}{c}
\Omega_1 \\
\vec{\theta}_1 \\
\Phi_1
\end{array}
\right] \underset{\mathcal{D}_2}{\longmapsto} \ldots
\underset{\mathcal{D}_m}{\longmapsto} \left[
\begin{array}{c}
\Omega_m \\
\vec{\theta}_m \\
\Phi_m
\end{array}
\right]
\end{eqnarray}
where each bracket represents, with the set
$[\Omega,\vec{\theta},\Phi]$ the effect of the transformation.
Define now the symmetric matrix $B_m$ taken from $\Omega$ as:
\begin{eqnarray}
B_m=\left[
\begin{array}{ccc}
\omega_{11} &\ldots  &\omega_{1m}\\
\vdots & & \vdots \\
\omega_{m1} &\ldots  &\omega_{mm}
\end{array}
\right]
\end{eqnarray}

\begin{eqnarray}
\Phi_m=\frac{\sqrt{i}^m}{\sqrt{\det B_m}}e^{-i\pi.
{^t}(\theta_1,\ldots,\theta_m) B_m^{-1}(\theta_1,\ldots,\theta_m)
}
\end{eqnarray}
with for $1\leq i\leq m$ and $1\leq j\leq m$
\begin{eqnarray}
(\omega_{m+1})_{i,j}=\frac{(-1)^{i+j}}{\det B_m} \det \left[
\begin{array}{cccccc}
\omega_{11} &\ldots  &\omega_{1,j-1} &\omega_{1,j+1} &\ldots  &\omega_{1,m}\\
\vdots & & \vdots &\vdots & & \vdots \\
\omega_{i-1,1} &\ldots  &\omega_{i-1,j-1} &\omega_{i-1,j+1} &\ldots  &\omega_{i-1,m}\\
\omega_{i+1,1} &\ldots  &\omega_{i+1,j-1} &\omega_{i+1,j+1} &\ldots  &\omega_{i+1,m}\\
\vdots & & \vdots &\vdots & & \vdots \\
\omega_{m,1} &\ldots  &\omega_{m,j-1} &\omega_{m,j+1} &\ldots
&\omega_{m,m}
\end{array}
\right] \nonumber \\
\end{eqnarray}
that is essentially the (i,j) cofactor of  $B_m$ . In a more compact way we can
write
\begin{eqnarray}
\Omega_{m+1}\vert_{(i,j)=1\ldots,m}=-B_m^{-1}
\end{eqnarray}
for $m<i$ and $1\leq j\leq m$
\begin{eqnarray}
(\omega_{m+1})_{i,j}=\frac{1}{\det B_m} \det \left[
\begin{array}{ccccccc}
\omega_{11} &\ldots  &\omega_{1,j-1} & \omega_{1i} &\omega_{1,j+1} &\ldots  &\omega_{1,m}\\
\vdots & & \vdots &\vdots  &\vdots & & \vdots \\
\omega_{m,1} &\ldots  &\omega_{m,j-1} & \omega_{mi}
&\omega_{m,j+1} &\ldots  &\omega_{m,m}
\end{array}
\right] \nonumber \\
\end{eqnarray}
This is obtained by substitution of the j\textsuperscript{th}
column by the i\textsuperscript{th} line. Hence if $m<i$ and
$m<j$, we have:
\begin{eqnarray}
(\omega_{m+1})_{i,j}=\frac{1}{\det B_m} \det \left[
\begin{array}{cccc}
\omega_{11} &\ldots   &\omega_{1,m} &\omega_{i,j}\\
\vdots & & \vdots  &\vdots \\
\omega_{m1} &\ldots   &\omega_{m,m} &\omega_{m,j}\\
\omega_{i1} &\ldots   &\omega_{i,m} &\omega_{i,j}\\
\end{array}
\right]
\end{eqnarray}
The right hand side is a $(m+1)\times (m+1)$ matrix obtained from
$B_m$ having a piece of the i\textsuperscript{th} line and
j\textsuperscript{th} column with $m<i$ and $m<j$. Finally
\begin{eqnarray}
\left[
\begin{array}{c}
\theta_1^m\\
\vdots\\
\theta_m^m\\
\end{array}
\right] =B_m^{-1} \left[
\begin{array}{c}
\theta_1\\
\vdots\\
\theta_m\\
\end{array}
\right]
\end{eqnarray}
and for all $i > m$
\begin{eqnarray}
\theta_i^m=\theta_i+\sum_{j=1}^m(\omega_{m+1})_{ij}\theta_j
\end{eqnarray}
These results may be proven by recursion. The proof is uniquely
based on the expansion of the determinant into minors. For example
for $(\omega_{m+1})_{11}$, one has to compute:
\begin{eqnarray}
(\omega_m)_{11}-\frac{(\omega_m)_{1m}^2}{(\omega_m)_{mm}}
\end{eqnarray}
with
\begin{itemize}
\item
\begin{eqnarray}
(\omega_m)_{11}=\frac{(\tilde{B}_m)_{11}}{\det B_m}
\end{eqnarray}
where $\tilde{B}_m$ is  the matrix formed by cofactors of $B_m$
\item
\begin{eqnarray}
(\omega_m)_{mm}=\frac{\det B_{m+1}}{\det B_m}
\end{eqnarray}
with $\det B_{m+1}= \omega_{m+1,m+1} \det B_m +\sum_{i=1}^m
\sum_{j=1}^m \omega_{i,m+1}\omega_{m+1,j} (\tilde{B}_m)_{i,j}$
\item
\begin{eqnarray}
(\omega_m)_{im}=\sum_{i=1}^m \omega_{i,m+1}
\frac{(\tilde{B}_m)_{i,1}}{\det B_m}
\end{eqnarray}
\end{itemize}
Hence
\begin{eqnarray}
&& (\omega_m)_{11}-\frac{(\omega_m)_{1m}^2}{(\omega_m)_{mm}} = \frac{\omega_{m+1,m+1}(\tilde{B}_m)_{11}}{\det B_{m+1}}\\
&& +\frac{1}{\det B_{m+1} \det B_{m}}\sum_{i=1}^m \sum_{j=1}^m
\omega_{i,m+1}\omega_{m+1,j}
\left((\tilde{B}_m)_{11}(\tilde{B}_m)_{ij}-(\tilde{B}_m)_{1j}(\tilde{B}_m)_{i1}
\right) \nonumber
\end{eqnarray}
Knowing that the determinant is invariant (up to a sign) under
permutations of lines or columns, it is sufficient to compute:
\begin{eqnarray}
&& (\tilde{B}_m)_{11}( \tilde{B}_m)_{mm}-(\tilde{B}_m)_{1m}(\tilde{B}_m)_{m1} = \left( \sum_{\sigma \in \mathcal{S}_{m-1}} (-1)^{\epsilon(\sigma)}\prod_{i=1}^{m-1} \omega_{i+1,\sigma (i)+1}  \right) \nonumber\\
& &\left( \sum_{\sigma^{\prime} \in \mathcal{S}_{m-1}} (-1)^{\epsilon(\sigma^{\prime})}\prod_{i=1}^{m-1} \omega_{i,\sigma^{\prime} (i)}  \right)-\left( \sum_{\sigma \in \mathcal{S}_{m-1}} (-1)^{\epsilon(\sigma^{\prime})}\prod_{i=1}^{m-1} \omega_{i,\sigma (i)+1} \right) \nonumber\\
&& \left( \sum_{\sigma^{\prime} \in \mathcal{S}_{m-1}}
(-1)^{\epsilon(\sigma)}\prod_{i=1}^{m-1}
\omega_{i+1,\sigma^{\prime} (i)}  \right)
\end{eqnarray}
Where $\mathcal{S}_{m-1}$ is the symmetric group of permutations
of order $(m-1)$. After  expansion and regrouping, we get the
double sum:
\begin{eqnarray}
 \sum_{\sigma \in \mathcal{S}_{m-1}} \sum_{\sigma^{\prime} \in \mathcal{S}_{m-1}  }
 (-1)^{\epsilon(\sigma)+\epsilon(\sigma^{\prime}) }\prod_{i=1}^{m-1} \prod_{j=1}^{m-1}
 \left( \omega_{i+1,\sigma(i)+1}\omega_{j,\sigma(j)}-\omega_{i,\sigma(i)+1}\omega_{j+1,
 \sigma(j)}  \right) \nonumber\\.
\end{eqnarray}
Now the product of permutations $\sigma \otimes \sigma^{\prime}
\in \mathcal{S}_{m-1}\times \mathcal{S}_{m-1}$ may be represented
as $\sigma \otimes \sigma^{\prime}  \in \mathcal{S}_{m}\times
\mathcal{S}_{m-2}$  and the previous double sum is recast under
the form:
\begin{eqnarray}
& &  \sum_{\sigma \in \mathcal{S}_{m}} \sum_{\sigma^{\prime} \in \mathcal{S}_{m-2}  }(-1)^{\epsilon(\sigma)+\epsilon(\sigma^{\prime}) }\prod_{i=1}^{m} \prod_{j=1}^{m-2} \left( \omega_{i,\sigma(i)}\omega_{j+1,\sigma(j)+1} \right) \\
& = & \left( \sum_{\sigma \in \mathcal{S}_{m}}
(-1)^{\epsilon(\sigma)}\prod_{i=1}^{m\omega_{i,\sigma(i)}}\right)
\left( \sum_{\sigma \in \mathcal{S}_{m-2}
}(-1)^{\epsilon(\sigma)}\prod_{i=1}^{m-2}\omega_{i+1,\sigma(i)+1}
\right) \nonumber\\
& = & \det B_m \det \left[
\begin{array}{ccc}
\omega_{12} &\ldots &\omega_{1,m-1}\\
\vdots & &\vdots\\
\omega_{m-1,2} &\ldots &\omega_{m-1,m-1}.
\end{array}
\right] \nonumber
\end{eqnarray}
For opposite permutations of lines and columns we have the following
relation:

\begin{eqnarray}
 &&(\tilde{B}_m)_{ij}( \tilde{B}_m)_{kk^{\prime}}-(\tilde{B}_m)_{ik}(\tilde{B}_m)_{jk^{\prime}} =\\
 &&\det B_m \det
\left[
\begin{array}{cccccccc}
\omega_{11} &\ldots &\hat{\omega}_{1,j} &\ldots &\hat{\omega}_{1,k^{\prime}} &\dots &\omega_{1,m}\\
\vdots & &\vdots & &\vdots & &\vdots\\
\hat{\omega}_{i1} &\ldots &\hat{\omega}_{i,j} &\ldots &\hat{\omega}_{i,k^{\prime}} &\dots &\hat{\omega}_{i,m}\\
\vdots & &\vdots & &\vdots & &\vdots\\
\hat{\omega}_{k1} &\ldots &\hat{\omega}_{k,j} &\ldots &\hat{\omega}_{k,k^{\prime}} &\dots &\hat{\omega}_{k,m}\\
\vdots & &\vdots & &\vdots & &\vdots\\
\omega_{m1} &\ldots &\hat{\omega}_{m,j} &\ldots &\hat{\omega}_{m,k^{\prime}} &\dots &\omega_{m,m}\\
\end{array}
\right]\nonumber\\
\end{eqnarray}
The last matrix is of size $(m-2)\times (m-2)$ obtained from $B_m$
by removal of columns $j$ and $k^{\prime}$ and of the lines
 $i$ and $k$ . Consequently:

\begin{eqnarray}
& & (\omega_m)_{11}-\frac{(\omega_m)_{1m}^2}{(\omega_m)_{mm}} = \frac{  ( \tilde{B}_m)_{11} \omega_{m+1,m+1}}{\det B_{m+1}} + \left(  \sum_ {i=1}^m\sum_ {j=1}^m
\frac{\omega_{i,m+1}\omega_{j,m+1}}{\det B_{m+1}} \det \right.  \nonumber \\
& &\left.
\left[
\begin{array}{cccccccc}
\omega_{11} &\ldots &\hat{\omega}_{1,j} &\ldots &\hat{\omega}_{1,k^{\prime}} &\dots &\omega_{1,m}\\
\vdots & &\vdots & &\vdots & &\vdots\\
\hat{\omega}_{i1} &\ldots &\hat{\omega}_{i,j} &\ldots &\hat{\omega}_{i,k^{\prime}} &\dots &\hat{\omega}_{i,m}\\
\vdots & &\vdots & &\vdots & &\vdots\\
\hat{\omega}_{k1} &\ldots &\hat{\omega}_{k,j} &\ldots &\hat{\omega}_{k,k^{\prime}} &\dots &\hat{\omega}_{k,m}\\
\vdots & &\vdots & &\vdots & &\vdots\\
\omega_{m1} &\ldots &\hat{\omega}_{m,j} &\ldots
&\hat{\omega}_{m,k^{\prime}} &\dots &\omega_{m,m}\\
\end{array}
\right] \right).
\end{eqnarray}

But this expression corresponds to the expansion of the determinant of
\begin{eqnarray}
(\omega_m)_{11}= \frac{(\tilde{B}_{m+1})_{11} }{\det B_{m+1}}=\det
\left[
\begin{array}{ccc}
\omega_{22} &\ldots &\omega_{2,m+1}\\
\vdots & &\vdots\\
\omega_{m+1,2} &\ldots &\omega_{m+1,m+1}
\end{array}
\right]
\end{eqnarray}
with respect to line $(m+1)$ and column $(m+1)$. Other results may
be proved along the same line, i.e. using the expansion of the
determinant into minors. Hence if $\Omega$ is invertible, one can
perform the duality transformation with respect to all indices
$k_j$, and the constants have the expressions:
\begin{eqnarray}
\Phi_d & = & \frac{\sqrt{i}^d}{\sqrt{\det \Omega}} e^{-i {^t}\vec{\theta} \Omega^{-1} \vec{\theta}} \nonumber\\
 \Omega_d & = & - \Omega^{-1}\nonumber\\
\vec{\theta}_d & = & \Omega^{-1} \vec{\theta}
\end{eqnarray}
Each sum $\sum_{k_{i}}$ is replaced by
$\sum_{k_{i}}+\mathcal{O}_i(1)$, as seen before. Proceeding as for
$d=2$, one can extend the two-dimensional result to the set of
invertible matrices with non-integer diagonal elements. Due to
linearity, this formula remains valid for sums on a bounded domain
$\mathcal{D} \in \Z^d$. The hypothesis of a bounded domain is
necessary since otherwise the Gaussian sums are not converging.
This is not the case for theta functions. Consequently, the result
is also valid for matrices $\Omega \in M_d(]0,1[)_{inv}$.

There remains the extension to $\R \setminus \Z$ to be examined. For this, let us
introduce $L_{ij} \in M_d(\{0,1\})$ such that
$(L_{ij})_{k,k^{\prime}}=0$ for all $k,k^{\prime}$ except for
$(k=i,k^{\prime}=j)$ or  $(k=j,k^{\prime}=i)$. Compute now:
\begin{eqnarray}
^d\sigma_{\mathcal{D}}^{[\vec{a},\vec{b}]} (\Omega+L_{ij},\vec{\theta}) & = & \sum_{\vec{k} \in \mathcal{D}} e^{i\pi \left( ^t(\vec{n}-\frac{\vec{a} }{2}) (\Omega+L_{ij}) (\vec{n}-\frac{\vec{a} }{2})   +2   {^t}(\vec{n}-\frac{\vec{a} }{2})(\vec{\theta} -\frac{\vec{b} }{2})\right)} \\
 &= & \sum_{\vec{k} \in \mathcal{D}} e^{i\pi \left( ^t(\vec{n}-\frac{\vec{a} }{2}) \Omega (\vec{n}-\frac{\vec{a} }{2})+2 {^t}(\vec{n}-\frac{\vec{a} }{2})(\vec{\theta} -\frac{\vec{b} }{2})  + {^t}(\vec{n}-\frac{\vec{a} }{2}) L_{ij} (\vec{n}-\frac{\vec{a} }{2}) \right)} \nonumber \\
&=& i^{a_ia_j}  \sum_{\vec{k} \in \mathcal{D}} e^{i\pi \left(
^t(\vec{n}-\frac{\vec{a} }{2}) \Omega (\vec{n}-\frac{\vec{a}
}{2})+2 {^t}(\vec{n}-\frac{\vec{a} }{2})(\vec{\theta}
-\frac{\vec{b} }{2})  -(a_in_j+a_jn_i) \right)} \nonumber
\end{eqnarray}
since:
\begin{eqnarray}
{^t}(\vec{n}-\frac{\vec{a} }{2}) L_{ij} (\vec{n}-\frac{\vec{a} }{2})=2(n_i-\frac{a_i}{2})(n_j-\frac{a_j}{2})=2n_in_j-(a_in_j+a_jn_i)+\frac{a_ia_j}{2}\nonumber\\
\end{eqnarray}
We set $\vec{b^{\prime} }\equiv \vec{b}+a_i\vec{e_j}+a_j\vec{e_i}
\mod (1,1)$ and obtain:
\begin{eqnarray}
^d\sigma_{\mathcal{D}}^{[\vec{a},\vec{b}]} (\Omega+L_{ij},\vec{\theta}) & = & (-i)^{a_ia_j}  \sum_{\vec{k} \in \mathcal{D}} e^{i\pi \left( ^t(\vec{n}-\frac{\vec{a} }{2}) \Omega (\vec{n}-\frac{\vec{a} }{2})+2 {^t}(\vec{n}-\frac{\vec{a} }{2})(\vec{\theta} -\frac{\vec{b^{\prime}} }{2})   \right)} \nonumber \\
&= &  (-i)^{a_ia_j}
{^d}\sigma_{\mathcal{D}}^{[\vec{a},\vec{b^{\prime}}]}
(\Omega,\vec{\theta})
\end{eqnarray}
Now let $L_{i} \in M_d(\{0,1\})_{sym}$ and
$(L_i)_{k,k^{\prime}}=0$ except for $k=i$, $k^{\prime}=1$. Then:
\begin{eqnarray}
^d\sigma_{\mathcal{D}}^{[\vec{a},\vec{b}]} (\Omega+L_{i},\vec{\theta}) =(-1)^{(1-a_i)n_i}(\sqrt{i})^{a_i} \sum_{\vec{k} \in \mathcal{D}} e^{i\pi \left( ^t(\vec{n}-\frac{\vec{a} }{2}) (\Omega) (\vec{n}-\frac{\vec{a} }{2})   +2   {^t}(\vec{n}-\frac{\vec{a} }{2})(\vec{\theta} -\frac{\vec{b} }{2})\right)} \nonumber \\
\end{eqnarray}
but with $\vec{b^{\prime} }= \vec{b}+(1-a_i)\vec{e_j}$, we have
the result
\begin{eqnarray}
^d\sigma_{\mathcal{D}}^{[\vec{a},\vec{b}]}
(\Omega+L_i,\vec{\theta}) =(\sqrt{i})^{a_i}
{^d}\sigma_{\mathcal{D}}^{[\vec{a},\vec{b^{\prime}}]}
(\Omega,\vec{\theta}),
\end{eqnarray}

and may state the theorem:
\begin{theor}
\label{theo6}
With the convention $\sqrt{-1}=i$ and $\forall \Omega \in
M_d^{\ast}([0,1[)_{inv}$ and $\forall \vec{\theta}\in ]0,1[^d$,
the Gaussian sum in $d$ dimensions admits the duality formula:
\begin{itemize}
\item
\begin{eqnarray}
^d\sigma_{\mathcal{D}}^{[\vec{a},\vec{b}]} (\Omega,\vec{\theta}) =\frac{(\sqrt{i})^d}{\det \Omega}
 e^{-i\pi {^t}\vec{\theta} \Omega^{-1} \vec{\theta}} \sum_{0\leq i_1,\ldots,i_d \leq d}\mathcal{O}(1) ^{i_1+\ldots+i_d}
{^d}\sigma_{\mathcal{P}_1^{i_1}\ldots\mathcal{P}_d^{i_d}\Omega\mathcal{D}}^{[\vec{a},\vec{b}]} (-\Omega^{-1},\Omega^{-1}\vec{\theta})
\nonumber\\
\end{eqnarray}
We use also the convention $\mathcal{P}^0=\mathbb{I}d$, the
identity map.
\item For $L_{ij} \in M_d(\{0,1\})_{sym}$ introduced before and $\vec{b^{\prime} }\equiv \vec{b}+a_i\vec{e_j}+a_j\vec{e_i} \mod (1,1)$ we also have:
\begin{eqnarray}
^d\sigma_{\mathcal{D}}^{[\vec{a},\vec{b}]}
(\Omega+L_{ij},\vec{\theta}) = (-i)^{a_ia_j}
{^d}\sigma_{\mathcal{D}}^{[\vec{a},\vec{b^{\prime}}]}
(\Omega,\vec{\theta})
\end{eqnarray}
\item And for $L_{i} \in M_d(\{0,1\})_{sym}$ introduced before and $\vec{b^{\prime} }= \vec{b}+(1-a_i)\vec{e_j}$, we obtain:
\begin{eqnarray}
^d\sigma_{\mathcal{D}}^{[\vec{a},\vec{b}]}
(\Omega+L_i,\vec{\theta}) =(\sqrt{i})^{a_i}
{^d}\sigma_{\mathcal{D}}^{[\vec{a},\vec{b^{\prime}}]}
(\Omega,\vec{\theta})
\end{eqnarray}
\end{itemize}
The functions $^dC_{\mathcal{D}}^{[\vec{a},\vec{b}]}
(\Omega,\vec{\theta}) $ and $^dS_{\mathcal{D}}^{[\vec{a},\vec{b}]}
(\Omega,\vec{\theta}) $ verify similar duality relations.
\end{theor}

If $\Omega$ is not invertible, one may perform a partial duality
transformation. Up to a permutation of indices, the duality
transformation may be applied to those of the indices for which
the minors are non-zero and the number of indices corresponds
simply to the rank of the matrix $\Omega$. For the remaining
indices the summation can be performed, since they are geometrical
sums.

\section[Applications]{Applications to the computation of $\xi_n[\beta,x;P_i]$}
\setcounter{equation}{0}
Now back to Eq. \ref{eq3304} which we shall write using the notation of the previous section.
The matrix $\Omega$ belongs to $M^{\ast}_{2q-1}(\R)_{sym}$ and has
the form
\begin{eqnarray}
\Omega= \frac{\beta}{\sum P_i} \left[
\begin{array}{cccc}
P_2(\sum P_i-P_2) & -P_2P_3 &\ldots &-P_2P_{2q}\\
-P_2P_3 &P_3(\sum P_i-P_3) &\vdots &-P_3P_{2q}\\
\vdots &\vdots &\ddots &\vdots \\
-P_2P_{2q} & -P_3P_{2q} &\ldots &P_{2q}(\sum P_i-P_{2q})
\end{array}
\right]\nonumber \\
\end{eqnarray}
we note $\sum$ for $\sum_{i=2}^{2q}$ and
\begin{eqnarray}
\vec{\theta}=x \left[
\begin{array}{c}
P_2\\
P_3\\
\vdots\\
P_{2q}
\end{array}
\right] \in Vect_{2q-1}(\R)
\end{eqnarray}
It is easy to see that $\det \Omega =0$ ($ \mbox{rank} (\Omega)\leq
2(q-1)$). By just adding all the lines, we get a line of $0$. Let
us compute the minor:
\begin{eqnarray}
& & \det\frac{\beta}{\sum P_i} \left[
\begin{array}{cccc}
P_2(\sum P_i-P_2) & -P_2P_3 &\ldots &-P_2P_{2q-1}\\
-P_2P_3 &P_3(\sum P_i-P_3) &\vdots &-P_3P_{2q-1}\\
\vdots &\vdots &\ddots &\vdots \\
-P_2P_{2q-1} & -P_3P_{2q-1} &\ldots &P_{2q-1}(\sum P_i-P_{2q-1})
\end{array}
\right]\nonumber \\
& & =\beta^{2(q-1)} \frac{P_2\ldots P_{2q-1}}{(\sum P_i)^{2(q-1)}}  \det \left[
\begin{array}{cccc}
\sum P_i-P_2& -P_2 &\ldots &-P_2\\
-P_3 &\sum P_i-P_3 &\vdots &-P_3\\
\vdots &\vdots &\ddots &\vdots \\
-P_{2q-1} & -P_{2q-1} &\ldots &\sum P_i-P_{2q-1}
\end{array}
\right]\nonumber\\
\end{eqnarray}

Now replacing the 1\textsuperscript{st} line by the sum of all the
lines, we obtain:
\begin{eqnarray}
 = \frac{\beta^{2(q-1)}P_2\ldots P_{2q}}{(\sum P_i)^{2(q-1)}} \det \left[
\begin{array}{cccc}
1 &1&\ldots &1\\
-P_3 &\sum P_i-P_{3} &\vdots &-P_3\\
\vdots &\vdots &\ddots &\vdots \\
-P_{2q-1} & -P_{2q-1} &\ldots &\sum P_i-P_{2q-1}\end{array}
\right] \nonumber\\
\end{eqnarray}
Replacing the $j$\textsuperscript{th} column by the difference of
the $j$\textsuperscript{th} and first column. Doing it for all
$j\geq2$, we are led to the expression:
\begin{eqnarray}
 && = \beta^{2(q-1)}\frac{P_2\ldots P_{2q}}{(\sum P_i)^{2(q-1)}} \det \left[
\begin{array}{ccccc }
1 &0&0 &\ldots &0\\
-P_3 &\sum P_i &0 &\vdots &0\\
-P_4 &0 &\sum P_i &\vdots &0\\
\vdots &\vdots &\vdots &\ddots  &\vdots \\
-P_{2q-1} & 0 &0 &\ldots &\sum P_i
\end{array}
\right] \nonumber\\
&=& \beta^{2(q-1)}\frac{P_2\ldots P_{2q}}{(\sum P_i)}
\not=0
\end{eqnarray}
Consequently we have the lemma
\begin{lemma}
\label{lem6}
\begin{eqnarray}
\mbox{rank} (\Omega)=2(q-1)
\end{eqnarray}
\end{lemma}

This result implies the duality transformation of $\xi_n$ only with
respect to indices $k_2,\ldots,k_{2q-1}$. Thus we have:
\begin{eqnarray}
\Phi_{2(q-1)}=\left(
\frac{i}{\beta}\right)^{(q-1)}\sqrt{\frac{\sum P_i}{P_2\ldots
P_{2q}}}e^{-i\pi {^t}(\theta_2,\ldots,
\theta_{2q-1})\Omega^{-1}_{2(q-1)} (\theta_2,\ldots,
\theta_{2q-1})}
\end{eqnarray}
with
\begin{eqnarray}
\Omega\vert_{i,j=2,\ldots,2q-1}\overset{def}{=}  \bar{\Omega}
=\frac{1}{\sum P_i}\left[
\begin{array}{ccc }
P_2(\sum P_i-P_2) &\ldots &-P_2P_{2q-1}\\
\vdots  &\ddots  &\vdots \\
-P_2P_{2q-1} &\ldots &P_{2q-1}(\sum P_i-P_{2q-1})\\
\end{array}
\right] \nonumber\\
\end{eqnarray}
Using the results of the above section we also obtain $\forall
i=2,\ldots, 2q-1$
\begin{eqnarray}
& & \omega_{2q-1}\vert_{i,2q}=\frac{1}{(\sum P_i)^{2q-3}P_2\ldots P_{2q}}\\
& &\det \left[
 \begin{array}{ccccc}
P_2(\sum P_i-P_2)&\ldots &-P_2P_{2q}  &\ldots &-P_2P_{2q-1}\\
-P_2P_3   &\ldots &-P_3P_{2q} &\ldots  &-P_2P_{2q-1}\\
\vdots     &\ddots  &\vdots &  &\vdots \\
-P_2P_i    &\ldots &-P_{2q}P_i &\ldots  &-P_iP_{2q-1}\\
\vdots    & &\vdots &\ddots    &\vdots \\
-P_2P_{2q-1}  &\ldots &-P_{2q-1}P_{2q} &\ldots  &-P_{2q-1}(\sum P_i-P_{2q-1})\\
\end{array}
\right] \nonumber
\end{eqnarray}
In the last matrix we substitute the i\textsuperscript{th} column
by the sum of all the columns of the matrix such that
i\textsuperscript{th} column becomes:
\begin{eqnarray}
[ -P_iP_2, \ldots,P_i(P_1+\ldots +\hat{P_i} + \ldots P_{2q}) ,
\ldots, -P_iP_{2q-1} ]
\end{eqnarray}
Such that $  \omega_{2q-1}\vert_{i,2q}=1$.We also have:
\begin{eqnarray}
 \left[
 \begin{array}{c}
\theta^{2q-1}_2 \\
\vdots\\
\theta^{2q-1}_{2q-1} \\
\end{array}
\right]  & = & \bar{\Omega}^{-1}
 \left[
 \begin{array}{c}
\theta_2 \\
\vdots\\
\theta_{2q-1} \\
\end{array}
\right]
\nonumber \\
\theta^{2q}_{2q-1} &= &\theta_{2q}+ \ldots+ \theta_2
\end{eqnarray}
At the end we get the form of $\Omega_{2q-2}$:
\begin{eqnarray}
\Omega_{2q-2}=
 \left[
 \begin{array}{cc}
\left(
 \begin{array}{ccc}
& & \\
&\bar{\Omega}^{-1} &\\
& &
\end{array}
\right)   &
 \begin{array}{c}
1 \\
\vdots\\
1
\end{array}\\
 \begin{array}{ccc}
1 &\ldots & 1
\end{array}
&0
\end{array}
\right]
\end{eqnarray}
where $\bar{\Omega}$ is the $(2q-2) \times (2q-2)$ matrix defined
above. So considering all the results modulo  $1$, the matrix
$\Omega_{2q-2}$ does not admit coupling between $k_{2q}$ and $k_i$
for $i=2,\ldots,2q-1$. Consequently with
$$A_{\vec{n}}=A_{\underbrace{n-1,\ldots,n-1}_{2q-2 \, times}}$$
and $\theta_i=xP_i$, $\forall i=2,\ldots,2q-1$
\begin{eqnarray}
&& \mathcal{D}_{2q-1} \circ \dots \circ  \mathcal{D}_{2}  \left( \xi_n(\beta,x;P_2,\ldots,P_{2q})\right)
= \left( \frac{i}{\beta} \right)^{q-1}\sqrt{\frac{\sum_{i=1}^{2q}P_i}{P_2\ldots P_{2q}}} \\
& &e^{-i\pi {^t} (\theta_2,\ldots,\theta_{2q-1})\bar{\Omega}^{-1} (\theta_2,\ldots,\theta_{2q-1})}
\sum_{k_1=0}^{n-1}\sum_{k_{2q}=0}^{n-1} e^{2\pi x (k_{2q}-k_1) \sum P_i  }  \sum_{2\leq i_2,\ldots,i_{2q-1} \leq 2q-1}\nonumber\\
& &  \mathcal{O}(1)^{i_2+\ldots+i_{2q-1}}.{^{2q-2}}\sigma_{\mathcal{P}_{i_2} \circ \ldots \circ
\mathcal{P}_{i_{2q-1}[\bar{\Omega}A_{\vec{n}}]}}^{[\vec{0},\vec{0}]}\left(-\frac{\bar{\Omega}^{-1}}{\beta},
\frac{\bar{\Omega}^{-1}}{\beta}\vec{\bar{\theta}} \right) \nonumber
\end{eqnarray}
where  $\vec{\bar{\theta}}$ stands for the vector:
\begin{eqnarray}
\vec{\bar{\theta}}=\left[\begin{array}{c}\theta_2 \\ \vdots\\
\theta_{2q-1} \end{array} \right]
\end{eqnarray}
Hence:
\begin{eqnarray}
& & \xi_n(\beta,x;P_2,\ldots,P_{2q})=\mathcal{O}(1)\sqrt{\frac{\sum_{i=1}^{2q}P_i}{P_2\ldots P_{2q}}} \sum_{k_1}^{n-1}\sum_{k_{2q}}^{n-1} e^{2\pi x (k_{2q}-k_1)\sum P_i}
\nonumber  \\
& & \sum_{2\leq i_2,\ldots,i_{2q-1} \leq 2q-1}
\sigma_{\mathcal{P}_{i_2} \circ \ldots \circ
\mathcal{P}_{i_{2q-1}[\bar{\Omega}A_{\vec{n}}]}}^{[\vec{0},\vec{0}]}\left(-\bar{\Omega}^{-1},
\bar{\Omega}^{-1}  \vec{\bar{\theta}} \right)
\end{eqnarray}
Here we are reduced to compute the $\sigma$-functions, with
$\Omega$ a matrix with integer entries of the form:
\begin{eqnarray}
\sigma_{A_{\vec{n}}}^{[\vec{0},\vec{0}]}\left(\lambda\Omega,\vec{\theta}\right)=\sum_{\vec{k}\in
A_{\vec{n}}}e^{i\pi (\lambda{^t}\vec{k} \Omega \vec{k} +
{^t}\vec{k}\vec{\theta})}
\end{eqnarray}
where $\lambda \in \R$
\begin{lemma}
\label{lem6b} For any matrix $\Omega$ with integer entries, there
exists a basis $(\vec{k}^{\prime})$ such that the quadratic form
$^t\vec{k}\Omega\vec{k}$ becomes:
\begin{eqnarray}
{^t}\vec{k} \Omega \vec{k}=\lambda_1{k^{\prime}_1}^2+\ldots +
\lambda_{2q-2}{k^{\prime}_{2q-2}}^2={^t}\vec{k}^{\prime} \Lambda
\vec{k}^{\prime}
\end{eqnarray}
where $\Lambda$ is a diagonal matrix with rational entries. If we
note $\vec{k}^{\prime}=U\vec{k}$, $U$ has also integer entries.
\end{lemma}
\proof Let us write down explicitly the expression of the
quadratic form:
\begin{eqnarray}
\sum_{ij}\omega_{i,j}k_ik_j &= & \sum_{1\leq i \leq d}\omega_{ii}k^2_i+2\sum_{1\leq i<j\leq d}
\omega_{ij}k_ik_j \nonumber\\
& = &\omega_{mm}\left( k_m+\sum_{i=1,\, i\not=m}^d\frac{\omega_{mi}}{\omega_{mm}}k_i\right)^
2-\omega_{mm}\left(\sum_{i=1,\, i\not=m}^d\frac{\omega_{mi}}{\omega_{mm}}k_i\right)^2 \nonumber\\
&+ & 2\sum_{1\leq i<j\leq d,\, (i,j)\not=m } \omega_{ij}k_ik_j \nonumber\\
& = &\omega_{m}{k^{\prime}}^2_m + \sum_{i=1,\,
i\not=m}\omega^{\prime}_{ii}{k^{\prime}}_i^2+2\sum_{1\leq i<j\leq
d,\, (i,j)\not=m }
\omega^{\prime}_{ij}{k^{\prime}}_i{k^{\prime}}_j
\end{eqnarray}
So by redefining the $\omega^\prime_{ij}$ and proceeding by
induction we obtained the expected result.
 $\blacksquare$

The transition matrix $U$ used for the
change of basis can be taken with integer entries instead of
rational entries. Thus $\forall i=1,\ldots,d$,
$k^{\prime}_i=\sum_{j=1}^dB_{ij}k_j$ and $U$ is of course
invertible. Moreover, up to a permutation of indices one can
insure that the next minor is non-zero. We set then
\begin{eqnarray}
U^{\prime}= \left[
 \begin{array}{ccc}
U_{11}&\ldots & U_{1,d-1}\\
\vdots & &\vdots \\
U_{d-1,1}&\ldots & U_{d-1,d-1}
\end{array}
\right]
\end{eqnarray}
and $\det(U^{\prime})\not=0$. Let $\mathcal{D}$ be a bounded
domain of $\N^d$ and let $\vec{P} \in U( \mathcal{D})$. Then
$\vec{P}$, by construction, is a vector with integer components
and $\exists \vec{k} \in \mathcal{D}$ such that $\vec{P} =
U\vec{k}$. At this stage, one should show that $U(\mathcal{D})$
can be decomposed into integer sub-lattices $R_i$ of lattice
spacing $\delta_i \in \N$. Thus $\vec{P} \in R_i$, then
$\vec{P}+\delta_i \vec{e}_j \in R_i$ with $\vec{e}_j$,
$j=1,\ldots, d$ is a unit vector. $R_i$ is defined inside the
boundary of $U(\mathcal{D})$. To prove this last property one must
solve in $\N^d$, the system $\forall i=1,\ldots,d-1$:
\begin{eqnarray}
\sum_{j=1}^dB_{ij}k_j &= &0 \nonumber \\
\sum_{j=1}^dB_{dj}k_j &= &\delta
\end{eqnarray}
We have thus
\begin{eqnarray}
U^{\prime}=\left[
 \begin{array}{c}
k_1\\
\vdots \\
k_{d-1}
\end{array}
\right] =-k_d \left[
 \begin{array}{c}
U_{1,d}\\
\vdots \\
U_{1,d-1}
\end{array}
\right]
\end{eqnarray}
and as $\det U^{\prime}\not=0$ then:
\begin{eqnarray}
\left[
 \begin{array}{c}
k_1\\
\vdots \\
k_{d-1}
\end{array}
\right] =-\frac{k_d}{\det U^{\prime}} \tilde{U}^{\prime} \left[
 \begin{array}{c}
U_{1,d}\\
\vdots \\
U_{1,d-1}
\end{array}
\right]
\end{eqnarray}
where $\tilde{U}^{\prime}$ is a matrix of cofactor of $U^{\prime}$
(with integer entries). Let us take $k_d=-\det U^{\prime}$ so we
have determined in $\N^{d}$ the solution the above equations :
\begin{eqnarray}
\left[
 \begin{array}{c}
k_1\\
\vdots \\
k_{d-1}
\end{array}
\right] =\tilde{U}^{\prime} \left[
 \begin{array}{c}
U_{1,d}\\
\vdots \\
U_{1,d-1}
\end{array}
\right]
\end{eqnarray}
It is clear that any multiple of this vector is a solution. Let us
compute now $\delta$, defined by:
\begin{eqnarray}
\delta=\sum_{j=1}^d U_{dj}k_j=U_{dd}\det
U^{\prime}+\sum_{i=1}^{d-1} \sum_{j=1}^{d-1} U_{dj}
\tilde{U}^{\prime}_{ij} U_{id}
\end{eqnarray}
Thus $\delta=\det U$ by expanding with respect to the last line
and column, we obtain:

\begin{proposition}
\label{prop2} Let $U(\mathcal{D})$ be a bounded domain of $\N^d$
and $U$ an integer invertible matrix . Then there exists at most
$\delta^d$ hypercubess $R_i \in \N^d$ of lattice spacing $\delta_i$
($\delta_i$ being a divisor of $\delta$, with  $\delta= \det U$),
such that $\forall P\in U(\mathcal{D})$, $\exists  1\leq i \leq
\delta^d$ for which $P\in R_i$. The hypercubes
$(R_i)_{i=1,\ldots,\delta^d}$ are all limited by boundaries of
$U(\mathcal{D})$.
\end{proposition}

\begin{proposition}
\label{prop3} Let $(\mathcal{F}_n)_{n\in \N}$ be a family of
hypercubes of $\N^d$, defined by $\vec{k}=(k_1,\ldots,k_d) \in
\mathcal{F}_n$ with $\forall 1\leq i \leq d$, $k_i \in \N$ and
$1\leq k_i\leq n$. Let $(\mathcal{D}_n)_{n\in \N}$ be an
increasing sequence of connected domains of $\N^d$ such that
there exists $t \in \R$ for which $\forall n\in \N$:
$\mathcal{F}_n\subset \mathcal{D}_n \subset \mathcal{F}_{tn}$.
Then, uniformly with respect to $\theta$:
\begin{eqnarray}
^d\sigma_{\mathcal{D}_n}^{[\vec{a},\vec{b}]}(\Omega,\vec{\theta})\underset{n\to \infty}{\sim}{^d}\sigma_{\mathcal{F}_n}^{[\vec{0},\vec{0}]}(\Omega,\vec{\theta})
\end{eqnarray}
\end{proposition}

\proof

$\N^d$ as a partially ordered set, we can easily prove that
$\forall (\vec{a},\vec{b}) \in Vect_d(\{0,1\})$ , $\forall \Omega
\in M_d(\R)$ and $\forall \vec{\theta} \in Vect_d(\R)$ the mapping
${^d}\sigma_{(.)}^{[\vec{a},\vec{b}]}(\Omega,\vec{\theta})$
defined as follows is increasing:
\begin{eqnarray}
\begin{array}{cccc}
{^d}\sigma_{(.)}^{[\vec{a},\vec{b}]}(\Omega,\vec{\theta}): & \mathcal{P}(\N^d) & \longrightarrow &\mathbb{C} \\
&\mathcal{F}& \longmapsto & {^d}\sigma_{\mathcal{F}}^{[\vec{a},\vec{b}]}(\Omega,\vec{\theta})
\end{array}
\end{eqnarray}
So for $n$ sufficiently large we have:
${^d}\sigma_{\mathcal{F}_n}^{[\vec{a},\vec{b}]}(\Omega,\vec{\theta})
\leq {^d}
\sigma_{\mathcal{D}_n}^{[\vec{a},\vec{b}]}(\Omega,\vec{\theta})
\leq {^d}
\sigma_{\mathcal{F}_{tn}}^{[\vec{a},\vec{b}]}(\Omega,\vec{\theta})$.
Now taking in account that:
${^d}\sigma_{\mathcal{F}_{tn}}^{[\vec{a},\vec{b}]}(\Omega,\vec{\theta})=\mathcal{O}({^d}\sigma_{\mathcal{F}_{n}}^{[\vec{a},\vec{b}]}(\Omega,\vec{\theta}))$,
 we can conclude that
${^d}\sigma_{\mathcal{D}_n}^{[\vec{a},\vec{b}]}(\Omega,\vec{\theta})=\mathcal{O}({^d}\sigma_{\mathcal{F}_{n}}^{[\vec{a},\vec{b}]}(\Omega,\vec{\theta}))$
and is homogeneous in $\vec{\theta}$. $\blacksquare$

\begin{remark}
\label{remm}
\end{remark}
Here we just recall  the result of [DM] that: $\forall \chi \in
[1/2,1]$ $\exists x\in \R$ such that
$^1\sigma^{[0,0]}_n(x,0)=\mathcal{O}(n^\chi)$. $\\ \\$ According
to [HL pp. 202], the hypothesis of uniformity is automatically
verified in Gaussian sums with $d=1$. This hypothesis remains
valid for $d$-dimensional Gaussian sums. Let
$R_1(\vec{n}),\dots,R_d(\vec{n})$ the sub-lattices making up
$B(\tilde{\Omega}A_{\vec{n}})$. Then we have:
\begin{eqnarray}
&& \sum_{0\leq i_2,\ldots,i_{2q-2} \leq 2q-1}\mathcal{O}(1)^{i_2+\ldots+i_{2q-1}}. {^{2q-2}}\sigma_{\mathcal{P}_{i_{2}}  \circ \ldots\circ \mathcal{P}_{i_{2q-1}}   (\tilde{\Omega}A_{\vec{n}})        }^{[\vec{0},\vec{0}]}(-\frac{\bar{\Omega}^{-1}}{\beta},\frac{\bar{\Omega}^{-1}}{\beta}\vec{\bar{\theta}}) \nonumber\\
&&=\sum_{j=1}^s \sum_{\overset{ i_2 \in [1,2q-1],\ldots,}{\underset{i_{2q-2} \in [1,2q-1]}{}}}
\mathcal{O}(1)^{i_2+\ldots+i_{2q-1}}. {^{2q-2}}\sigma_{\mathcal{P}_{i_{2}}  \circ \ldots\circ
\mathcal{P}_{i_{2q-1}}   (R_j)    
}^{[\vec{0},\vec{0}]}(-\frac{\Lambda}{\beta}^{-1},-\frac{\Lambda}{\beta}^{-1}\vec{\bar{\theta}}^{\prime}) \nonumber\\
\end{eqnarray}
with $\vec{\bar{\theta}}^{\prime}=B^{-1}\vec{\bar{\theta}}$,
where $B$ is the matrix  of basis change described in lemma
\ref{lem6b} such that $\Lambda$ is a diagonal matrix with integer
(or rational) entries.  Since 
$\Lambda$ has integer matrix elements, it follows from the above formula that there exists a finite
number of sub-lattices for which the multi-variable sum is reduced
to a product of ($2q-2$) factors of one-dimensional Gaussian
sums where each factor is of the form ${^{1}}\sigma_{ 
(R_j)    
}^{[\vec{0},\vec{0}]}(-\frac{1}{\lambda\beta},-\frac{1}{\lambda\beta}{\theta})
$ where $\lambda$ is an eigenvalue of $\Lambda$. The following section will
consider the problem of the asymtotic behavior of such Gaussian sums.

\subsection{Estimations of Gaussian sums using the duality formula}
\setcounter{equation}{0}

We shall now present the theorem which motivates the previous
construction. Let us recall the method of Hardy et Littelwood  using the
duality formula for the computation of the Gaussian sums. They
give the expression, $\forall x\in ]0,1[$, $\forall \theta \in
]0,1[$ and $\forall (a,b) \in \{0,1\}$, and any $n\in \N$:
\begin{eqnarray}
\label{2eq5b}
^1\sigma_n^{[a,b]}(x,\theta) =\mathcal{O}(1) \sqrt{nxx_1\ldots x_\nu}+\frac{\mathcal{O}(1)}{\sqrt{xx_1\ldots x_\nu}}
\end{eqnarray}

Recall briefly the steps:
For convenience let us denote $x_0=x$. $\forall i=0,\ldots, \nu$,
we have $0<x_i<1$. $x_i$ are the rest of the development in
continuous fractions of $x$:
\begin{eqnarray}
x=\frac{1}{a_1+\frac{1}{a_2+\ldots}}\overset{not.}{=}[a_1,a_2,\ldots,]
\end{eqnarray}
and $\forall i>0$ the $(x_i)_{i\in \N}$ are given inductively by
the relation:
\begin{eqnarray}
x_{i-1}= \frac{1}{a_i+x_i}
\end{eqnarray}
Hence, for any $n$, there is  an integer $\nu$  determined in such a way that :
\begin{eqnarray}
\label{eq3614}
nx_0\ldots x_{\nu} \leq 1\leq nx_0\ldots x_{\nu-1}
\end{eqnarray}
Through this condition, the first term on the r.h.s. in Eq.6.1.1 becomes
irrelevant with respect to the second one. Let us put $\nu=\nu(n)$
in the above expression. So there exists a positive constant $H$
such that:
\begin{eqnarray}
\label{eq3615}
x_0\ldots x_{\nu} \geq H a^{-1}_{\nu(n)} x_0\ldots x_{\nu-1}
\end{eqnarray}
Hence we obtain by using Eq.\ref{eq3614} and Eq.\ref{eq3615}:
\begin{eqnarray}
\frac{1}{\sqrt{x_0\ldots x_{\nu}}} =\mathcal{O}( \sqrt{a_{\nu(n)} n})
\end{eqnarray}
If we use the trivial estimation $(x_ix_{i+1}<\frac{1}{2})$ we
deduce that necessarily $\forall \epsilon>0$:
\begin{eqnarray}
\nu(n) < \frac{2+\epsilon}{\ln 2}\ln{n}
\end{eqnarray}
thus:
\begin{itemize}
\item if $a_n=\mathcal{O}(1)$ then $^1\sigma_n^{[a,b]}(x,\theta)=\mathcal{O}(\sqrt{n})$
\item if $a_n=\mathcal{O}(n^\rho)$ then $^1\sigma_n^{[a,b]}(x,\theta)=\mathcal{O}(\sqrt{n}(\ln n)^{\frac{\rho}{2}})$
\item if $a_n=\mathcal{O}(e^{n\rho})$, with $0<\rho<\frac{\ln2}{2}$ then $^1\sigma_n^{[a,b]}(x,\theta)=\mathcal{O}(\sqrt{n}n^{\frac{\rho}{\ln 2}+\epsilon})$
\end{itemize}
Nevertheless let us recall that the equidistribution of
$e^{i\pi xk^2}$, for any irrational $x$, [K] and [KN] give the upper bound of:
\begin{eqnarray}
^1\sigma_n^{[a,b]}(x,\theta)=o(n)
\end{eqnarray}
Now we must use other results of number theory concerning the
computation of continuous fraction expansion of a fraction of the numbers occurring as
arguments in our Gaussian sums ${^{1}}\sigma_{ 
(R_j)    
}^{[\vec{0},\vec{0}]}(-\frac{1}{\lambda\beta},-\frac{1}{\lambda\beta}{\theta})
$ , of the form:  \begin{eqnarray}
\label{2eq8} ^1\sigma_n^{[a,b]}\left( \frac{ux+v}{tx+w},\theta
\right) \leq K.\left( ^1\sigma_n^{[0,0]}(x,0) \right) 
\end{eqnarray}
where $ (u,v,t,w)\in \N^4$.

 We
will refer to the article 
of Raney [R] and to the survey of Van der Poorten [VdP]. Let us
introduce some notations:
\begin{eqnarray}
\begin{array}{cccc}
\Psi_k: &]0,1[ &\longrightarrow &{\N^*}^{\N} \\
&x &\longmapsto &[a_1,a_2,\ldots,a_k]
\end{array}
\end{eqnarray}
such that
\begin{eqnarray}
\Psi_k(x)=\frac{1}{a_1+\frac{1}{a_2+\frac{1}{\ldots+\frac{1}{a_k}}}}
\end{eqnarray}
Denote $\Psi_ \infty$ the map which associates to any real number
its continuous fraction. $\Psi_\infty$ is a homeomorphism from
$]0,1[$ to ${\N^*}^{\N}$. Referring to the theorem of convergence
for continuous fraction, for any $x\in ]0,1[$ we have:
\begin{eqnarray}
\Psi^{-1}_\infty \circ \Psi_k(x) \underset{ k \to \infty}{\to} x
\end{eqnarray}
Consider now the matrix representation $\{R,L\}$ of the expansion
in continuous fractions of $x$:
\begin{eqnarray}
R= \left[
\begin{array}{cc}
1 &1 \\ 0 &1
\end{array}
\right]
\qquad
L= \left[
\begin{array}{cc}
1 &0 \\ 1&1
\end{array}
\right]
\end{eqnarray}
Here $\forall k\in \N^*$ one gets
\begin{eqnarray}
R^k= \left[
\begin{array}{cc}
1 &k \\ 0 &1
\end{array}
\right]
\qquad
L^k= \left[
\begin{array}{cc}
1 &0 \\ k &1
\end{array}
\right]
\end{eqnarray}
So  $\Psi_k^{-1}([a_1,\ldots,a_k]) \in \Q$ is determined by  the formula:

\begin{eqnarray}
\Psi_k^{-1}([a_1,\ldots,a_k])=\frac{\sum_{i=1}^2(R^{a_1}L^{a_2}\ldots R^{a_k})_{1,i}}{\sum_{i=1}^2(R^{a_1}L^{a_2}\ldots R^{a_k})_{2,i}}
\end{eqnarray}
Now we use the main result of the articles quoted above:

The continuous fraction of $\frac{ax+b}{cx+d}$ is related to that
of $x$ if $ad-ac\not=0$. To this end, let us consider the set of
 $2\times 2$ matrices of positive integer entries:

\begin{eqnarray}
\mathcal{E}_n = \left \lbrace M= \left( \begin{array}{cc} a &b \\c &d \end{array}\right)   
\in M_2(\N) | \quad n = \det
M\not=0, a>c, b<d \right \rbrace
\end{eqnarray}
The number of such matrices is finite [R]. Denote by
$\mathcal{E}_n=\{A_1,\ldots, A_N\}$. The commutation relations
obey  some restrictions: $\forall i=1,\ldots,N$, $\exists
i^{'},i^" =1,\ldots,N$ such that:
\begin{eqnarray}
A_iR^{k^1_R}L^{k^1_L} & = & L^{k^{'1}_L}R^{k^{'1}_R}A_{i^{'}} \nonumber \\
A_iL^{k^2_L}R^{k^2_R} & = & R^{k^{'2}_R}L^{k^{'2}_L}A_{i^{''}}
\end{eqnarray}
with $ ( k^1_ R,k^1_L)$ positive exponents such that $1\leq k^1_
R+k^1_L \leq n$ and so on. The algorithm which allows to compute, for a given n, the matrices in the left hand side is given
below. We note that these commutation relations are invariant under transposition:
\begin{eqnarray}
A_iR^{k^1_R}L^{k^1_L} & = & L^{k^{1^{'}}_L}R^{k^{1^{'}}_R}A_{i^{'}} \nonumber\\
^tA_{i^{'}} R^{k^{1^{'}}_R} L^{k^{1^{'}}_L}& = & L^{k^1_L}R^{k^1_R} {^t}A_i
\end{eqnarray}
Note that for any $n\in \N^*$
\begin{eqnarray}
\left( \begin{array}{cc} n &0\\0 &1 \end{array}\right) R=R^n \left( \begin{array}{cc} n &0\\0 &1 \end{array}\right) \nonumber\\
\left( \begin{array}{cc} n &0\\0 &1 \end{array}\right) L=L^n \left( \begin{array}{cc} n &0\\0 &1 \end{array}\right)
\end{eqnarray}
This provides an algorithm to compute the expansion into continued fraction of the values of the function $x\mapsto
\frac{ax+b}{cx+d}$
\subsection{Illustration of the algorithm}
\setcounter{equation}{0}

We now shall illustrate this method for $n=2$. The set
$\mathcal{E}_2$ is made up two matrices which represent the
multiplication and the division by $2$:
\begin{eqnarray}
\mathcal{E}_2 & = & \{A,A^\prime\} \\
A  =  \left\lbrack \begin{array}{cc} 2 &0 \\ 0 &1  \end{array}  \right\rbrack & & A^\prime  =  \left\lbrack \begin{array}{cc} 1 &0 \\ 0 &2  \end{array}  \right\rbrack
\end{eqnarray}
The rules of commutation are compiled in the following tabulation:
\begin{eqnarray}
\mbox{
\begin{tabular}[b]{|c|c|c|}
\hline
 &$A$ &$A^\prime$ \\
\hline
\hline
$A$&$R:R^2$ & $LR:RL$ \\
& $L^2:L$ & \\
\hline
$A^\prime $&$RL:LR$ & $R^2:R$ \\
 & &$L:L^2$ \\
\hline
\end{tabular}
}
\end{eqnarray}

The table is to be red following the exemple of the first entry:$ AR = R^2A$. These relations are equivalent to the
following distinct cases:
\begin{eqnarray}
\begin{array}{ccc}
x=[0,2a,b,\ldots] &\Rightarrow &2x=[0,a,2b,\ldots] \\
x=[0,2a+1,b,\ldots] &\Rightarrow &2x=[0,a,1,1,(b-1)/2,\ldots] \\
\end{array}
\label{2eq6}
\end{eqnarray}
The entire determination of $2x$ from $x$ is done by induction
with respect to those rules. For example, if $x=[7,2,5,\ldots]$
then
\begin{eqnarray}
\label{2eq7}
2x=[3,1,1,0,1,1,2,\ldots]=[3,1,2,1,2,\ldots]
\end{eqnarray}

Let us introduce some notations: $[x_0,x_1,\ldots]$ with
$0<x_i<1$, $\forall i \in \N$, the sequence of the rest deduced by
the decomposition into continuous fractions of $x=x_0$. We shall
denote the resulting sequence of the rest of $2x$ by
$2x=x_0^\prime=[x_0^\prime,x_1^\prime,\ldots]$. Thus we have , by
definition, the  relation:
\begin{eqnarray}
x_i=-b_{i+1}+\frac{1}{x_{i+1}}
\end{eqnarray}
where $x=[0;b_1,b_2,\ldots]$. As the formula Eq \ref{2eq6} shows,
the processus of commutation (resulting of the multiplication by
$2$ generates a lag length between the two sequences
($x=[0;b_1,b_2,\ldots]$ and $2x=[0;b_1^\prime
,b_2^\prime,\ldots]$). We shall define a function (depending on
the decomposition of $x$) which measures the length of $2x$ with
respect to $x$.
\begin{eqnarray}
\begin{array}{cccc}
k_x^\prime: & \N &\longrightarrow & \N \\
 &k &\longmapsto &k_x^\prime(k)
\end{array}
\end{eqnarray}
such that :
\begin{itemize}
\item if $2$ divides $b_1$ ( in what follows denoted as: $2\mid b_1$ ),  then $k_x^\prime(k_1)=k_1$ and
$k_x^\prime(k_2)=k_2$
\item if $2$ does not divide $b_1$ ( in what follows denoted as $2\nmid b_1$) then $k_x^\prime(k_1)=k_1+3$ and
\begin{itemize}
\item if   $2\nmid b_2$ then $k_x^\prime(k_2)=k_x^\prime(k_1)+1$
\item if $2\mid b_2$ then $k_x^\prime(k_2)=k_x^\prime(k_1)+3$
\end{itemize}
\end{itemize}
and so one... We recall the trivial  reduction which occurs when
one of the components of the decomposition is zero, we shall refer
to the example Eq \ref{2eq7}. Note for this $x$ ,
$k_x^\prime(3)=5$.
\begin{lemma}
\label{lem8}
$\forall p\in \N$, let  $[x_0,x_1,\ldots]$, $[x^\prime_0,x^\prime_1,\ldots]$ be the rests of the expansion of $x$ and
$x^\prime= px $ respectively, then $\forall k \in \N$ $\exists C \in \N$ which divides $k$ (possibly $C=1$). such that
\begin{eqnarray}
x^\prime_1\ldots x^\prime_{k_x^\prime(k)}=Cx_1\ldots x_k
\end{eqnarray}
where $[x_0,x_1,\ldots]$, $[x^\prime_0,x^\prime_1,\ldots]$ are
respectively the rest sequence of $x$ and $px$.
\end{lemma}

\proof

We shall reduce the proof to the simplest case $p=2$. The other
case is quite the same but it turns out to be more difficult
because of an involved  algebra. We prove the lemma distinguishing
all the possibilities:
\begin{itemize}
\item if  $2\mid b_1$ then $b^\prime_1=\frac{b_1}{2}$.
Compute now $x_0^\prime x_1^\prime=x^\prime_0\left(-b^\prime_1+\frac{1}{x^\prime_0} \right)=1-b_1x_0=x_0x_1$.
Consequently $x^\prime_1=\frac{x_1}{2}$. Since $2\mid b_1$ then $b^\prime_2=2b_2$ and $x^\prime_1 x^\prime_2=1-b^\prime_2 x^\prime_1=1-b_2x_1=x_1x_2$ and $x_2^\prime=2x_1$.

At this level we have two possibilities either $2 \mid b_3$ and
the procedure repeats itself identically or $2\nmid b_3$, in this
case we shall refer to the following
\item
if $2\nmid b_1$ then $b^\prime_1\frac{b_1-1}{2}$ and
$b_2^\prime=b_3^\prime=1$. Hence let us calculate $x_0^\prime
x_1^\prime x_2^\prime x_3^\prime$: $x_0^\prime x_1^\prime
x_2^\prime x_3^\prime=-(2b_1^\prime +1)
x_0^\prime+2=-2b_1x_0+2=2(1-b_1x_0)=2x_1x_0$ so $x _1=  x_1^\prime
x_2^\prime x_3^\prime$. We establish another relation $x_0^\prime
x_1^\prime x_2^\prime =(1+b^\prime_1)x^\prime_0+1$ then
$x_1^\prime x_2^\prime=1+b^\prime_1-\frac{1}{x^\prime_0}=
\frac{1-x_1}{2}$ ie $x_1+2x^\prime_1x^\prime_2=1$ and we deduce
that $(1-x_1)x^\prime_3=2x_1$

\item If $2\nmid b_2$ then $b^\prime_4=\frac{b_2-1}{2}$ and $b^\prime_5=2b_3$. Calculating $x^\prime_3 x^\prime_4 x^\prime_5=x^\prime_3(1+b_5^\prime b_4^\prime)-b^\prime_5$, so
$x^\prime_4 x^\prime_5=1+b_2b_3-\frac{b_3}{x_1}=x_2x_3$.
We also prove that $-b_5^\prime x_4^\prime+1=x_2x_3=-b_3x_2+1$ so $x^\prime_4=\frac{x_2}{2}$
and $x^\prime_5=2x_3$.

\item If $2 \mid b_2$ then $b_4^\prime=\frac{b_2-2}{2}$ end $b_5^\prime =b_6^\prime=1$. $x_3^\prime x_4^\prime x_5^\prime x_6^\prime=2-(b_2-1)x^\prime_3$. So using $2x_1=x^\prime_3(1-x_1)$ we get
$x_4^\prime x_5^\prime x_6^\prime=x_2$. we also get $2x_4^\prime
x_5^\prime+x_2=1$. So let us summarize these results in the
following tables:
\end{itemize}
\begin{eqnarray}
\mbox{
\begin{tabular}[b]{|c|c|c|c|c|c|}
\hline
$x=$ &$2 b_1$ &$b_2 $&$2b_3$ & &  \\
\hline
$x_0$ &$x_1$ &$x_2$ &$x_3$  & &  \\
\hline
$2x =$ &$b_1 $&$2b_2$ &$b_3$ & &  \\
\hline
$x^\prime_0 =2x_0$ &$x^\prime_1=\frac{x_1}{2} $&$x^\prime_2=2x_2$ &$x^\prime_3=\frac{x_3}{2}$ & &  \\
\hline
\hline
$x=$ &$2 b_1$ &$b_2 $&$2b_3+1$ & &\\
\hline
$x_0$ &$x_1$ &$x_2$ &$x_3$ & & \\
\hline
$2x=$ &$b_1 $&$2b_2$ &$b_3$&1 &1  \\
\hline
$x^\prime_0 =2x_0$ &$x^\prime_1=\frac{x_1}{2} $&$x^\prime_2=2x_2$ &$\begin{array}{c} x^\prime_3 \\ x_3=x^\prime_3 x^\prime_4 x^\prime_5  \\  x^\prime_3 x^\prime_4 +2x_3=1  \end{array}  $ & $x^\prime_4 $&$x^\prime_5$  \\
\hline
\end{tabular}
}\nonumber \\
\end{eqnarray}
and we also have:
\begin{eqnarray}
\mbox{
\begin{tabular}[b]{|c|c|c|c|c|c|c|}
\hline
$x=$ &$2 b_1+1$ & & &$2b_2+1$ & $b_3$ & \\
\hline
$x_0$ &$x_1$ & & &$x_2$ &$x_3$ &\\
\hline
$2x =$ &$b_1 $&1 &1 &$b_2$ &$2b_3$ &  \\
\hline
$x^\prime_0 =2x_0$ &$\begin{array}{c} x^\prime_1 \\ x_1=x^\prime_1 x^\prime_2 x^\prime_3 \\   x^\prime_1 x^\prime_2 +2x_1=1  \end{array}  $ &$ x^\prime_2 $ &$x^\prime_3$ &$\frac{x_2}{2}$ & $2x_3$ &   \\
\hline
\hline
$x=$ &$2 b_1+1$ & & &$2b_2+2$ &  & \\
\hline
$x_0$ &$x_1$ & & &$x_2$ & &\\
\hline
$2x =$ &$b_1 $&1 &1 &$b_2$ &1 &1  \\
\hline
$x^\prime_0 =2x_0$ &$\begin{array}{c} x^\prime_1 \\ x_1=x^\prime_1 x^\prime_2 x^\prime_3   \\  x^\prime_1 x^\prime_2 +2x_1=1  \end{array}  $ &$ x^\prime_2 $&$x_3^\prime$ &$\begin{array}{c} x^\prime_4 \\ x_2=x^\prime_4 x^\prime_5 x^\prime_6  \\  x^\prime_4 x^\prime_5 +2x_2=1  \end{array}$& $x^\prime_5$ &$x^\prime_6$   \\
\hline
\end{tabular}
}\nonumber \\
\end{eqnarray}
Thus we have : $\forall k \in \N$ $x_1\ldots x_k= x^\prime_1\ldots
x^\prime_{k^\prime(k)}$ or $2 x_1\ldots x_k= x^\prime_1\ldots
x^\prime_{k^\prime(k)}$. This complete the proof of the lemma.
$\blacksquare$ $\\$

More generically, let $A\in \mathcal{E}_n$ goes through the string
$R^{a_1}L^{a_2}R^{a_3}\ldots$, commuting with respect to the two
matrices $\{R,L\}$: $[A R^{a_1}L^{a_2}R^{a_3}\ldots R^{a_k}]
\mapsto [ R^{a^{\prime}_1}L^{a^{\prime}_2}R^{a^{\prime}_3}$
$\ldots R^{a^{\prime}_{k^{\prime}}}A^{\prime}]$ with
$A^{\prime}\in \mathcal{E}_n$.  This procedure gives an algorithm
for the computation of:
\begin{eqnarray}
\Psi_{k^{\prime}(k)} \left( \frac{ax+b}{cx+d} \right)= [a^{\prime}_1,a^{\prime}_2,a^{\prime}_3,\ldots,a^{\prime}_{k^{\prime}(k)}]
\end{eqnarray}
which only depends on $[a_1,a_2,\ldots,a_k]$. Now using the above
lemma in the formula Eq \ref{2eq5b}, we obtain the theorem:
\begin{theor}
\label{theo7}
$\forall (u,v,t,w)\in \N^4$, $\forall x\in \R \setminus \Q$, $\forall \theta \in [0,1[$ and $\forall (a,b)\in \{0,1\}$, then
\begin{eqnarray}
^1\sigma_n^{[a,b]}\left( \frac{ux+v}{tx+w},\theta \right) =\mathcal{O}\left( ^1\sigma_n^{[0,0]}(x,0) \right)
\end{eqnarray}
\end{theor}

By applying this theorem to the above expression one can deduce
that:
\begin{eqnarray}
\xi_n(x,\beta;P_2,\ldots,P_{2q}) & = & \sqrt{\frac{\sum P_i}{P_2\ldots P_{2q}}} \sum_{k_1=0}^{n-1}\sum_{k_{2q}=0}^{n-1} e^{2i\pi x (k_{2q}-k_1)(\sum P_i)} \nonumber\\
& \times & \mathcal{O}({^1}\sigma_{n}^{[a,b]}(\beta,0))^{2q-2}
\end{eqnarray}

\begin{remark}
\end{remark}

The estimation depends on $(u,v,t,w)$. It remains an
essential question: does there exist  a constant $K$ independant of $ (u,v,t,w)\in \N^4$ such that\begin{eqnarray}
\label{2eq8} ^1\sigma_n^{[a,b]}\left( \frac{ux+v}{tx+w},\theta
\right) \leq K.\left( ^1\sigma_n^{[0,0]}(x,0) \right)
\end{eqnarray}
This result is not trivial to prove. In fact, if this assertion
were true, the characteristic function would be  analytical in a
neighborhood of zero. Consequently, this work only gives a partial
answer.

\begin{remark}
\label{rem4}
\end{remark}
We observe that this result is an optimal estimation [HL p225, theorem 
2.221] when applied to the matrix case: for any $\Lambda$ is an integer
invertible matrix, and for any $\beta\in \R \setminus \Q$, we have:
\begin{eqnarray}
{^d}\sigma_{A_{\vec{n}}}^{[\vec{a},\vec{b}]}(\beta \Lambda,\vec{\theta})=\mathcal{O}({^1}\sigma_{n}^{[0,0]}(\beta,0))^d
\end{eqnarray}
It cannot be replaced by a better estimate.

Now we can compute the mean value of $S_n^{2q}\left(\frac{\pi \beta}{4}\right)$ (see Eq. \ref{eq3305}):
\begin{eqnarray}
& & E\left[S_n^{2q}\left(\frac{\pi \beta}{4}\right) \right]=(-1)^q \left( \frac{4}{\pi} \right)^q \sum_{P_i \in \Zz} \frac{1}{ \prod P_i (\sum P_i)} \int_0^1 \xi_n(x,\beta;P_2,\ldots,P_{2q}) dx \nonumber\\
& & =(-1)^q \left( \frac{4}{\pi} \right)^q \sum_{P_i \in \Zz} \frac{\mathcal{O}({^1}\sigma_{n}^{[0,0]}(\beta,0))^{2(q-1)}}{( \prod P_i)^{\frac{3}{2}}\sqrt{\sum P_i}} \sum_{k_1=0}^{n-1}\sum_{k_{2q}=0}^{n-1}\int_0^1e^{2i\pi x (k_{2q}-k_1)\sum P_i}dx \nonumber\\
\end{eqnarray}
where $\prod$ stands for $\prod_{i=2}^{2q} $. Performing the
integral in $x$ yields the condition $k_1=k_{2q}$, we end up with:

\begin{eqnarray}
 & & E\left[S_n^{2q}\left(\frac{\pi \beta}{4}\right) \right]= n\sum_{P_i \in \Zz} 
\frac{\mathcal{O}({^1}\sigma_{n}^{[0,0]}(\beta,0))^{2(q-1)}}{( \prod P_i)^{\frac{3}{2}}\sqrt{\sum P_i}} \nonumber\\
& &=
n.c_q.
\mathcal{O}({^1}\sigma_{n}^{[0,0]}(\beta,0))^{2(q-1)}
\end{eqnarray}

$\mathcal{O}({^1}\sigma_{n}^{[0,0]}(x,0))^{2(q-1)}$ depends on $n$
but also on $(P_2,\ldots,P_{2(q-1)})$. Moreover $c_q$is finite because the moments
are perfectly defined (the integration is done over a compact set
$[0,2\pi[\times [0,2\pi[$). So:
\begin{eqnarray}
 E\left[S_n^{2q}\left(\frac{\pi \beta}{4}\right) \right]=n.\mathcal{O}\left({^1}\sigma_{n}^{[0,0]}\left(\beta,0\right)^{2(q-1)}\right)
\end{eqnarray}
\begin{remark}
\label{rem5}
\end{remark}
For $q=1$, we recover the exact result $
E\left[S_n^{2q}\left(\frac{\pi \beta}{4}\right) \right]=n$, which
is independent of the choice of $\beta$.

\begin{theor}
\label{theo8} The unique sequence $f_n$ ( up to an equivalence as $n\to \infty$) in order that:
\begin{eqnarray}
E\left[\frac{S_n\left(\frac{\pi \beta}{4}\right)}{f_n} \right] ^{2q}=\mathcal{O}(1)
\end{eqnarray}
with respect to $n$ for all q, is $f_n =\sqrt{n}$. This estimate is true  
for all $\beta$ such that ${^1}\sigma_{n}^{[0,0]}\left(\beta,0\right)=\mathcal{O}(\sqrt{n})$
\end{theor}

\proof

  We must choose a
$\beta$ having a bounded continuous fractions representation to have $f_n$ indenpendent of $q$. The proof results
straightforwardly from the remark \ref{remm} that is:
\begin{eqnarray}
\frac{f_n^2}{n^q}\left(\frac{{^1}\sigma_{n}^{[0,0]}\left(\frac{\pi \beta}{4},0\right)}{f_n}\right)^{2q}=\mathcal{O}(1)
\end{eqnarray}
$\blacksquare$
\begin{remark}
\label{rem6}
\end{remark}

Note that if $\beta \in \Q$ then [HL] ${^1}\sigma_{n}^{[0,0]}\left(\beta,0\right)=\mathcal{O}(n)$ and we recover the result established
earlier, i.e:
\begin{eqnarray}
 E\left[S_n^{2q}\left(\frac{\pi \beta}{4}\right) \right]=\mathcal{O} (n^{2q-1})
\end{eqnarray}
which means that there is no normalization of $S_n$ for which the
moments neither diverge nor be non-zero.

Examples of numbers $\beta$ which admit an expansion as an
infinite and bounded continued fractions are the quadratic irrationals. They have ultimately 
periodic continued fraction expansion. This class contains all 
the square roots of products of pairwise distinct prime integers,  and of
course we have, for those numbers, $\forall q\in \N$
\begin{eqnarray}
 E\left[S_n^{2q}\left(\frac{\pi \beta}{4}\right) \right]=\mathcal{O} (n^q)
\end{eqnarray}
Hence there exists here a normalization $\sqrt{n}$ of $S_n$ for
which all moments remain bounded with respect to $n$ and are not
zero.

\begin{rmq}
\label{rem7}
\end{rmq}
Any attempt to obtain information to the convergence  in  distribution  of $S_n/f_n$ by using numerical computations
cannot be correct since behaviors are radically different
according to the nature of the number $\beta$ and the
rational numbers are dense in $\R$. Finally all of these
calculations may be applied in the same way to any periodic
function instead of the signum function $\chi(y)$.

As to the problem of the convergence in distribution  of $S_n/\sqrt{n}$, we shall make some comments in the conclusion:

\section{Conclusions}
In this approach to determine the limiting distribution law of
$\frac{S_{n}}{\surd n}$ we have sought to compute the
asymptotic behavior of the moments  of
$\frac{S_{n}}{\surd n}$. If the parameter $\beta=\frac{\alpha}{4\pi}$ admits an
expansion in bounded continued fraction, the behavior obtained for
the expectation values of moments
$E\left(\frac{S_{n}^{2k}}{n^{k}}\right)= \mathcal O(1)\leq A_{k}$
may lead to the convergence of the series
$\sum_{k=0}^{\infty}\frac{(it)^{2k}}{2k!}A_{k}$, around $t=0$. In that case,  the
sequence $\Phi_{n}(t)$ converges towards $\Phi(t)$,
which is analytic near the origin. This implies the
existence of a limiting distribution with finite moments.
The estimation of the speed of the increase of the $A_k$ seems difficult and is still an open problem. It is thus difficult to have an idea of the limiting distribution without such estimation,
Although the procedure does not imply the convergence in distribution  for $\frac{S_{n}}{\surd  n}$, it shows that this
normalization leading to bounded second moment, fails to lead to bounded moments of higher orders when $\beta$ is an
irrational having no expansion in bounded continued fractions.

\bigskip

\noindent {\bf Acknowledgments}

It is a pleasure to thank Martine Queffelec for exchange of informations
about works on Gaussian sums and continued fraction
decomposition.

\newpage
\section*{ References}

[AR] V.\ Armitage. A.\ Rogers  \textit{"Gauss sums and quantum mechanics"}. J. Phys. A, \textbf{33}, 5993-6002, (2000).

[Bi] P.\ Billingsley \textit{" Convergence of Probability measures"}, Wisley, New York, (1978).

[BCS] L.A.\ Bunimovich, N.I.\ Chernov, Ya.\ Bunimovich,\textit{" Statistical properties of two-dimensional hyperbolic billards"}, Russ. Math. Survey, \textbf{46} , 47-106, (1991).

[BS]. L. A.\  Bunimovich, Ya.\ Sina\"{i}.  \textit{" Markov partitions for dispersed billiards"}, Comm. Math. Phys. \textbf{78}, no. 2, 247--280. (1980/81).

[CC]. J.D..\ Crawford, J.R.\ Cary \textit{" Decay of correlations in a chaotic measure preserving transformation"}. Physica D.  \textbf{6}, 2 23-232, (1983).

[CH1]. M.\ Courbage, D.\ Hamdan. \textit{'' Decay of Correlation and mixing properties in a dynamical system with zero entropy"}. Ergod. Th. Dynam. Syst.  \textbf{17}, no.1, 87-103,  (1997).

[CH2]. M.\ Courbage, D.\ Hamdan. \textit{" Chapman-Kolmogorov equation for non-markovian shift invariant measure"}. Ann. Prob. \textbf{22}, 1662-1677, (1994).

[CH3]  M.\ Courbage , D. \ Hamdan, {\it Unpredictability in some nonchaotic dynamical systems}, Phys.Rev.Lett. 74, 5166-5169, 1995.

[CFS]. I.P.\ Cornfel, S.V.\ Fomin, Ya.G.\ Sinai \textit{"Ergodic theory"}. Springer, New York, (1981).

[CYo]. N.I.\ Chernov, L.S.\ Young, ''\textit{Statistical properties of dynamical systems with some hyperbolicity}", Ann. of Math.  \textbf{147}, 585-650, (1998).

[DM]. F.M.\ Dekking, M.\ Mend\'{e}s France  \textit{"Uniform distribution modulo one: a geometrical viewpoint "}. J. Reine Angew. Math, \textbf{329}, 143-153 ,(1981).

[Go]. M.I.\ Gordin. \textit{"The Cental Limit Theorem for stationnary processes"}. (Russian) Dokl. Akad. Nauk SSSR, \textbf{188}, 739--741,
(1969).

[HL]. G.H.\ Hardy, J.E.\ Littlewood  \textit{"Problems of diophantine approximation II"}. Acta mathematica , \textbf{37}, 193-238,(1914).

[J]. S.\ Janson  \textit{"Some pairwise independant sequences for which the Central Limit Theorem fails "}. Stochastics, \textbf{23}, no 4, 439-448, (1988).

[K]. N.M.\ Korobov, H.\ Niederreiter. \textit{"Exponential sums and their applications"}. Mathematics and its applications (Kluwer academic plublishers) , Boston, (1992).

[KN]. L.\ Kuipers, H.\ Niederreiter. \textit{"Uniform distribution of sequences"}. Wiley, New York, (1994).

[L]. E.\  Lindel\"{o}f \textit{"Le calcul des r\'{e}sidus et ses applications \`a la th\'{e}orie des fonctions"}. Reprinted from the 1905 original by Les Grands Classiques Gauthier-Villars. Éditions Jacques Gabay, Sceaux, (1989).

[Li]. C.\ Liverani  \textit{"Central limit for deterministic systems. "} International Conference on Dynamical Systems (Montevideo, 1995), 56--75, Pitman Res. Notes Math. Ser., 362, Longman, Harlow, (1996).

[M]. D.\ Munford. \textit{"Tata lectures on theta functions "}. vol (1), Birkhauser, Boston, (1983).

[MMP]. R.S.\ Mackay, J.D\ Meiss, I.C.\ Percival \textit{"Transport in Hamiltonian systems"}, Physica D, no 1-2, \textbf{13}, 55-81, (1984).

[MOS]. W.\ Magnus. F.\ Oberhettinger, R.P.\ Soni \textit{"Formulas and theorems for special functions of mathematical physics"}, 3rd Ed. Springer. New York (1966).

[R]. N.\ Raney \textit{"On Continued Fractions and Finite Automata "}. Math. Ann, \textbf{206}, 265-283,(1973).

[SLB]. S.\ Seshadri. S.\ Lahshamibala. V.\ Balakrishnan \textit{"Control of wave packet revivals using geometric phases"}. J. Stat. Phys. \textbf{101}, 213-223, (2000).

[VdP]. A.J.\ Van der Porteen  \textit{"Diophantine analysis: roceeding of the Number theory section of the 1985 Australian Mathematical Society Convention  "}. New York: Cambridge University, (1985).

[Z]. Z.\ Zalcwasser  \textit{"Sur les polyn\^{o}mes associ\'{e}s aux fonctions modulaires Theta"}. Studia Mathematica \textbf{7}, 16-35, (1937).

[Za]. G.M.\ Zaslavsky, \textit{" Chaotic dynamics and the origin of statistical laws"}. Phys. Today, August. (1999),

\end{document}